\documentclass[journal]{IEEEtran}
\usepackage[table]{xcolor} 
\usepackage{amssymb}
\usepackage{stfloats}
\usepackage{graphicx}
\usepackage{multirow}
\usepackage{subfigure}
\usepackage{tabularx}
\usepackage{booktabs}
\usepackage{amsmath}
\usepackage{amsthm}
\usepackage{epsfig,epsf,color,balance,cite}
\usepackage{indentfirst}
\usepackage{mathrsfs}
\usepackage{dsfont}
\usepackage{utfsym}
\UseRawInputEncoding
\usepackage{easyReview}
\usepackage[justification=centering]{caption}
\usepackage[font=small,labelfont=md,labelsep=period]{caption}

\newtheorem{remark}{Remark}
\usepackage{algorithm, algorithmicx, algpseudocode}
\usepackage{enumerate}
\usepackage{makecell}
\usepackage{arydshln}
\newcounter{MYtempeqncnt}
\usepackage[colorlinks,bookmarksopen,bookmarksnumbered,citecolor=blue, linkcolor=blue, urlcolor=blue]{hyperref}

\makeatletter
\DeclareRobustCommand{\cev}[1]{%
	{\mathpalette\do@cev{#1}}%
}
\newcommand{\do@cev}[2]{%
	\vbox{\offinterlineskip
		\sbox\z@{$\m@th#1 x$}%
		\ialign{##\cr
			\hidewidth\reflectbox{$\m@th#1\vec{}\mkern4mu$}\hidewidth\cr
			\noalign{\kern-\ht\z@}
			$\m@th#1#2$\cr
		}%
	}%
}
\makeatother

\begin{document}
	\title{Revisiting XL-MIMO Channel Estimation: When Dual-Wideband Effects Meet Near Field}
	\vspace{-2em}
	\author{
		\IEEEauthorblockN{ 
			Anzheng Tang\IEEEauthorrefmark{1},
			Jun-Bo Wang\IEEEauthorrefmark{1},
			Yijin Pan\IEEEauthorrefmark{1}, 
			Tuo Wu\IEEEauthorrefmark{2},
			Yijian Chen\IEEEauthorrefmark{3},\\
			Hongkang Yu\IEEEauthorrefmark{3},
			and Maged Elkashlan\IEEEauthorrefmark{4}
		}
		\thanks{This work was supported in part by the National Natural Science Foundation of China under Grant No. 62371123, 62350710796 and 62331024, in part by ZTE Industry-University-Institute Cooperation Funds under Grant No. IA20240723011-PO0006, and in part by Research Fund of National Mobile Communications Research Laboratory Southeast University under Grant No. 2025A03. (\emph{Corresponding authors: Jun-Bo Wang and Yijin Pan})}
		\thanks{A. Tang, J.-B. Wang and Y. Pan are with the National Mobile Communications Research Laboratory, Southeast University, Nanjing 210096, China. (E-mail: \{anzhengt, jbwang, and panyj\}@seu.edu.cn). A. Tang is also with the Department of Electronic and Computer Engineering, Hong Kong University of Science and Technology, Hong Kong, China. (E-mail: \{eeaztang\}@ust.hk).
		}
		\thanks{T. Wu is with the School of Electrical and Electronic Engineering, Nanyang Technological University, 639798, Singapore (E-mail: tuo.wu@ntu.edu.sg).}
		\thanks{Y. Chen and H. Yu are with the Wireless Product Research and Development Institute, ZTE Corporation, Shenzhen 518057, China. (E-mail:\{yu.hongkang, chen.yijian\}@zte.com.cn)}
		\thanks{M. Elkashlan is with the School of Electronic Engineering and Computer Science, Queen Mary University of London, London E1 4NS, U.K. (E-mail: maged.elkashlan@qmul.ac.uk).}
	}
	\maketitle
	\vspace{-4em}
	\begin{abstract}
		The deployment of extremely large antenna arrays (ELAAs) in extremely large-scale multiple-input multiple-output (XL-MIMO) systems introduces significant near-field effects, such as spherical wavefront propagation and spatially non-stationary (SnS) properties. When combined with the dual-wideband effects inherent to wideband systems, these phenomena fundamentally alter the channel's sparsity patterns in the angular-delay domain, rendering existing estimation methods insufficient.
		To address these challenges, this paper reconsiders the channel estimation problem for wideband XL-MIMO systems. Leveraging the spatial-chirp property of array responses, we first quantitatively characterize the angular-delay domain sparsity of wideband XL-MIMO channels, revealing both global block sparsity and local common-delay sparsity. To effectively capture this structured sparsity, we then propose a novel column-wise hierarchical prior model that integrates a precision sharing mechanism and a Markov random field (MRF) structure. Building on this prior model, the channel estimation task is formulated as a multiple measurement vector (MMV)-based Bayesian inference problem.
		Tailored to the complex factor graph induced by this hierarchical prior, we develop a MMV-based hybrid message passing (MMV-HMP) algorithm. This algorithm performs message updates along the edges of the factor graph, and selectively applies either the variational message passing (VMP) or sum-product (SP) rules, depending on the factor-node structure and message tractability.
		Simulation results validate the effectiveness of the proposed column-wise hierarchical prior model through ablation studies and demonstrate that the MMV-HMP algorithm, while maintaining moderate computational complexity, consistently outperforms existing baselines which fail to capture the structured sparsity of wideband XL-MIMO channels.
	\end{abstract}  
	\begin{IEEEkeywords}
		Wideband XL-MIMO systems,  dual-wideband effects, near-field effects, channel estimation, message passing.    
	\end{IEEEkeywords}
	\IEEEpeerreviewmaketitle
	\vspace{-1em}
	\section{Introduction}
	\label{section1}
	{Wideband extremely large-scale multiple-input-multiple-
	output (XL-MIMO) has emerged as a pivotal technology to meet the capacity demands of future 6G communications \cite{Tutorial, WT1}. By exploiting the spatial degrees of freedom (DoFs) provided by extremely large antenna arrays (ELAAs), XL-MIMO significantly enhances both spectral and energy efficiencies\cite{XL_MIMO_T2, WT2, HOU2}. It also delivers substantial beamforming gains, effectively compensating the severe path loss of millimeter-wave (mmWave) frequencies. Additionally, the availability of GHz-wide bandwidth in the mmWave spectrum alleviates spectrum congestion, reinforcing the potential of mmWave XL-MIMO for next-generation wireless networks \cite{HOU1,WT3}.
	However, realizing these performance benefits critically depends on the accurate channel state information (CSI).}
	
	{Typically, wideband XL-MIMO systems adopt orthogonal frequency division multiplexing (OFDM) to combat the frequency-selective fading of wireless channels \cite{mmWavebandWidth}. 
	However, due to the pronounced impact of dual-wideband effects, it is challenging to acquire accurate CSI.
	Specifically, the large system bandwidth shortens the OFDM symbol duration, potentially causing the propagation delay difference across the antenna array to exceed the symbol duration, leading to the spatial wideband effect \cite{Dual_Wideband}. Concurrently, the varying central frequencies of subcarriers introduce the frequency-wideband effect, where phase shifts differ across subcarrier channels, further complicating the acquisition of CSI.}
	
	Moreover, with the deployment of ELAAs and operation at higher frequency bands, near-field effects, such as the spherical wavefront effect \cite{LoS_Angular, SnS_CE, NF3} and spatially non-stationary (SnS) properties \cite{SnS1, SnS2}, become prominent, exacerbating the dual-wideband effects. For instance, the sparsity structure of XL-MIMO channels in the angular-delay domain alters significantly. 
	Specifically, the curvature of the spherical wavefront means a single spatial frequency is no longer sufficient to characterize a propagation path;  instead, multiple spatial frequencies are required. This spatial frequency spread, combined with the frequency-wideband effect, results in a more pronounced angular spread compared to that of conventional massive MIMO systems \cite{Dual_Wideband, Dual_Wideband_2}. Additionally, the spherical wavefront effect disrupts the linear variations of path delay across antennas, while SnS properties influence both spatial frequency and delay spreads.
	These combined factors reshape the sparse patterns of XL-MIMO channels in the angular-delay domain, significantly complicating channel estimation in wideband XL-MIMO systems.
	\vspace{-1em}
	\subsection{Related Works}
	Due to the limited scattering resulting from the highly directional propagation behavior of mmWave, mmWave channels typically exhibit significant sparsity. Consequently, most existing channel estimation methods are designed to exploit this inherent sparsity structure to simplify and improve the estimation process. 
	For instance, leveraging the sparsity in the polar domain, various channel estimation and beam training schemes have been proposed to address the spherical effects in XL-MIMO channels \cite{NFBT2, NFCE2, NFCE4_2}. Considering the inherent spatial-domain sparsity induced by SnS properties, several Bayesian inference-based methods have been developed \cite{Bayesian1, Bayesian3} to mitigate the SnS effects. Additionally, exploiting the joint sparsity of XL-MIMO channels in both the spatial and angular domains, methods such as \cite{SnS_CE, TLGAMP} have proposed two-stage and joint visibility region (VR) detection and channel estimation algorithms, considering both spherical wavefront and SnS properties.
	However, despite these advancements, these methods predominantly focus on narrowband systems without the considerations of dual-wideband effects.
	
	While some studies have explored wideband channels, such works remain limited in scope. For instance, \cite{NFCE1} leverages the common polar-domain sparsity across subcarrier channels, employing the simultaneous orthogonal matching pursuit (SOMP) method for channel estimation. Nevertheless, as the bandwidth and array aperture increase in wideband XL-MIMO systems, the dual-wideband effects become more pronounced, challenging the assumption of common sparsity across subcarriers and diminishing the applicability of such methods.
	
	Considering the sparsity differences among subcarriers, \cite{Dual_Wideband_2, Gao_Jiabao_CE} proposed methods that exploit subcarrier-dependent variations in sparsity patterns. Specifically, \cite{Dual_Wideband_2} recognized the sparsity-dependent pattern across subcarriers in dual-wideband channels and proposed jointly utilizing information from multiple subcarriers to enhance dual-wideband channel estimation performance. \cite{Gao_Jiabao_CE} developed a deep learning (DL)-based approach by unfolding the sparse Bayesian learning (SBL) algorithm into a deep neural network (DNN), where each SBL layer is carefully designed to capture subcarrier-dependent sparsity through a tailored variance parameter update mechanism.
	In addition, \cite{FFBS_1, FFBS_2, FFBS_3} developed the super-resolution-based estimation techniques to enhance performance by exploiting angular-delay domain sparsity.  
	Unfortunately, these methods are tailored for far-field channels and cannot be readily extended to near-field scenarios due to the differences of sparsity patterns.
	
	More recently, methods addressing both the spherical wavefront effect and dual-wideband effects have been proposed. For example, \cite{NFCE4_4} introduced a bilinear pattern detection (BPD)-based approach to recover wideband XL-MIMO channels, while \cite{NFCE4_5} developed a message passing algorithm based on constrained Bethe free energy minimization. However, despite these advancements, none of these approaches explicitly incorporate SnS properties, highlighting a significant gap in wideband XL-MIMO channel estimation.
	\vspace{-1em}
	\subsection{Motivations and Contributions}
	{Due to the randomness of the environment and user locations, spherical wavefront effects and SnS properties are inevitable in wideband XL-MIMO systems. However, the channel estimation problem that jointly considers the spherical wavefront effect, SnS properties, and dual-wideband effects has not been well addressed to date. Specifically, a well-established channel model for wideband XL-MIMO systems is lacking, and the sparsity patterns of these channels in the angular-delay domain have yet to be fully explored. Additionally, with the increasing number of antennas and subcarriers in XL-MIMO systems, there is a pressing need for computationally efficient channel estimation algorithms to handle the growing complexity.}
	To address these issues, this paper investigates channel estimation for XL-MIMO systems by incorporating the spherical wavefront effect, SnS properties, and dual-wideband effects. The main contributions of this paper are summarized as follows:
	\begin{itemize}
		\item To accurately characterize the SnS dual-wideband XL-MIMO channels, we derive a spatial-frequency domain model that comprehensively captures the spherical wavefront propagation, SnS characteristics, and dual-wideband effects.
		By exploiting the spatial-chirp property of array responses, we quantitatively analyze the influence of spherical wavefront propagation, SnS properties, and dual-wideband effects on the angular and delay spreads.
		Furthermore, we reveal the resulting sparsity patterns in the angular-delay domain, which exhibit both global block sparsity and local common-delay sparsity.
		\item Exploiting the sparsity in the angular-delay domain, the channel estimation problem is formulated as a multiple measurement vector (MMV)-based Bayesian inference task.
		Recognizing the importance of an accurate prior model in channel reconstruction, we devise a novel column-wise hierarchical prior, which incorporates both a precision sharing mechanism and a Markov random field (MRF) structure. This design enables the model to effectively capture global block sparsity and local common-delay sparsity, thereby providing a more structured representation of the underlying channel characteristics.
		\item To achieve efficient Bayesian inference, resulting from the hierarchical prior, we propose a MMV-based hybrid message passing (MMV-HMP) algorithm that jointly leverages sum-product (SP) and variational message passing (VMP) rules.
		Thanks to this flexible message update strategy, the MMV-HMP algorithm can effectively manage the message updates for latent precision parameters with non-Gaussian distributions and accommodate the loops inherent in the complex factor graph, thereby enabling accurate inference.
		\item Compared to existing approaches, the proposed MMV-HMP algorithm achieves superior performance by jointly capturing the global block sparsity and local common-delay sparsity while maintaining moderate computational complexity. In addition, comprehensive ablation studies validate the effectiveness of the proposed column-wise hierarchical prior model, confirming its ability to capture the channel's structured sparsity.
	\end{itemize}
	\vspace{-1em}
	\subsection{Organization and Notations}
	The rest of the paper is organized as follows: 
	Section \ref{section2} introduces the mmWave XL-MIMO system model.
	Section~\ref{section3} develops a spatial-frequency channel model, incorporating near-field and dual-wideband effects, and analyzes the angular-delay characteristics of XL-MIMO channels.
	Section~\ref{section4} formulates the channel estimation problem and proposes a column-wise hierarchical sparse prior model.
	Section~\ref{section5} presents the MMV-HMP algorithm.
	Simulation results and conclusions are provided in Sections \ref{section6} and \ref{section7}, respectively.
	
	Notations: lower-case letters, bold-face lower-case letters, and bold-face upper-case letters are used for scalars, vectors and matrices, respectively; 
	The superscripts $\left(\cdot \right)^{\mathrm{T}}$ and $\left(\cdot \right)^{\mathrm{H}}$ stand for transpose and conjugate transpose, respectively;  
	$\otimes$ denotes the Kronecker product;
	$\mathbf{I}_{N}$ denotes a $N \times N$ identity matrix; 
	$\mathbb{C}^{M \times N}$ denotes a $M \times N$ complex matrix. 
	$\mathbf{1}_{N}$ denotes a $N \times 1$ all-one vector. 
	In addition, a random variable $x \in \mathbb{C}$ drawn from the complex Gaussian distribution with mean $x_0$ and variance $v$ is characterized by the probability density function $\mathcal{CN}(x;m,v) = \exp\{-{\left|x-m\right|^2}/{v}\}/{\pi v}$; a random variable $\gamma \in \mathbb{R}$ from Gamma distribution with mean $a/b$ and variance $a/b^2$ is characterized by the probability density function $\mathcal{G}a(\gamma; a, b) \propto \gamma^{a-1}\exp(-\gamma b)$.
	
	\begin{figure*}
		\centering
		\includegraphics[width=0.6\textwidth]{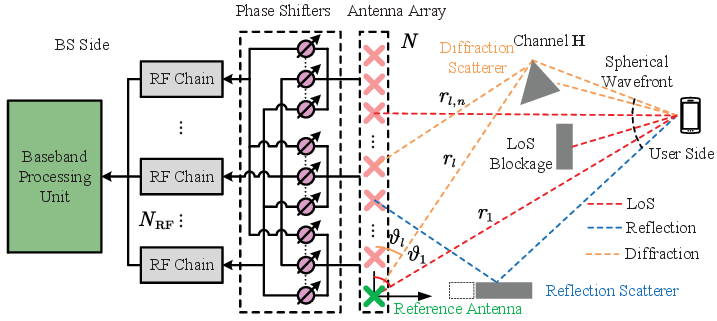}
		\caption{Illustration of the XL-MIMO system model.}
		\label{SystemModel}
	\end{figure*} 
	\section{System Model}
	\label{section2}
	Consider a mmWave XL-MIMO OFDM system, as shown in Fig.~\ref{SystemModel}, where a base station (BS) utilizes an ELAA comprising $N$ antennas arranged in a uniform linear configuration. 
	To reduce hardware cost and energy consumption, the BS adopts a hybrid analog-digital architecture, where $N_{\mathrm{RF}}$ radio frequency (RF) chains are connected to the ELAA through phase shifters. The spacing between adjacent antennas is $d = \lambda_c / 2$, where $\lambda_c = c / f_c$, with $c$ and $f_c$ indicating the speed of light and the central carrier frequency, respectively. 
	For the uplink channel estimation, we assume the $J$ users transmit mutual orthogonal pilot sequences to the BS \cite{PolarCS, FFBS_2, Pattern}, e.g., orthogonal time or frequency resources are utilized for different users to transmit pilot sequences. Therefore, channel estimation for each user is independent. Without loss of generality, we consider an arbitrary user and assume $K$ subcarriers are uniformly selected from the available subcarriers to carry pilots.
	
	In the uplink channel estimation phase, the BS combines the pilot signal using all RF chains associated with different beams. Let the total number of received beams be denoted by $M$, with $M$ being an integer multiple of $N_{\mathrm{RF}}$. Consequently, the BS requires $P = M / N_{\mathrm{RF}}$ channel uses to cycle through all beams for a given pilot symbol. Denote $\iota_{k,p}$ as the pilot symbol of the $k$-th subcarrier in the $p$-th channel use. Then, the received signal for the $k$-th subcarrier in the $p$-th channel use can be expressed as  
	\begin{equation}
		\mathbf{y}_{k,p} = \mathbf{W}_p \mathbf{h}_k \iota_{k,p} + \mathbf{n}_{k,p}, 
	\end{equation}
	where $\mathbf{W}_p \in \mathbb{C}^{N_{\mathrm{RF}}\times N}$ denotes the combining matrix in the $p$-th channel use, with each entry adhering to a constant modulus constraint and independently generated from the set $\{-1/\sqrt{N},1/\sqrt{N}\}$;
	$\mathbf{h}_k \in \mathbb{C}^{\mathrm{N}\times 1}$ denotes the spatial-domain channel vector for the $k$-th subcarrier, while $\mathbf{n}_{k,p} \sim \mathcal{CN}\left(\mathbf{n}_{k,p}; \mathbf{0},\beta^{-1}\mathbf{I}_{N_{\mathrm{RF}}}\right)$  represents the noise vector associated with the $k$-th subcarrier in the $p$-th channel use.
	
	Assuming that all-one pilot symbols are adopted, i.e., $\iota_{k,p} = 1$ for all $k$ and $p$, and collecting the pilot symbols across all $P$ channel uses, the received signal for the $k$-th subcarrier can be written in a compact form as
	\begin{equation}
		\mathbf{y}_{k} = \mathbf{W} \mathbf{h}_k + {\mathbf{n}}_k,
	\end{equation}
	where $\mathbf{W}  = [\mathbf{W}_1^{\mathrm{T}}, \mathbf{W}_2^{\mathrm{T}}, \cdots, \mathbf{W}_P^{\mathrm{T}}]^{\mathrm{T}} \in \mathbb{C}^{M\times N}$ and ${\mathbf{n}}_k = [\mathbf{n}_{k,1}^{\mathrm{T}}, \mathbf{n}_{k,2}^{\mathrm{T}}, \cdots, \mathbf{n}_{k,P}^{\mathrm{T}}]^{\mathrm{T}} \in \mathbb{C}^{M\times 1}$ are the collective receive combining matrix and the effective noise vector, respectively.
	Furthermore, concatenating the received signals corresponding to different pilot subcarriers, the overall received signal $\mathbf{Y} = \left[\mathbf{y}_1, \mathbf{y}_2, \cdots, 
	\mathbf{y}_{K}\right]\in \mathbb{C}^{M\times K}$ is given by
	\begin{equation}
		\mathbf{Y} = \mathbf{W}\mathbf{H} + {\mathbf{N}},
		\label{Y_pilot}
	\end{equation}
	where $\mathbf{H}=\left[\mathbf{h}_1, \cdots, \mathbf{h}_K\right] \in \mathbb{C}^{N\times K}$ and ${\mathbf{N}} = \left[{\mathbf{n}}_1, \cdots, {\mathbf{n}}_K\right] \in \mathbb{C}^{M\times K}$ indicating the spatial-frequency channel matrix and effective noise matrix, respectively.
	\vspace{-1em}
	\section{SnS Dual-Wideband XL-MIMO Channel Model}
	\label{section3}
	 In this section, we first derive a spatial-frequency channel model that incorporates the effects of spherical wavefront effect, SnS properties, and dual-wideband effects. Then, we quantitatively  analyze the angular-delay characteristics of SnS dual-wideband XL-MIMO channels.
	 \vspace{-1em}
	\subsection{Channel Modeling for SnS XL-MIMO Systems}
	In XL-MIMO systems, the use of ELAA and high-frequency bands facilitates near-field communications over hundreds of meters, invalidating the far-field plane wavefront assumption and necessitating the consideration of spherical wavefront effects. Additionally, SnS properties arise as different array portions observe varying propagation conditions, causing power variations across array elements. Furthermore, dual-wideband effects in wideband XL-MIMO systems influence channel characteristics, requiring careful modeling.
	
	Assume that there are $L$ propagation paths between the BS and user. Denote $\mathcal{L} = \left\{1,2,\cdots, L\right\}$ as the set of all propagation paths, where $l = 1$ refers to the line-of-sight (LoS) path and $l \ge 2$ indicates reflection or diffraction path, as illustrated  in Fig.~\ref{SystemModel}.
	Denote $\tau_{l,n}$ as the time delay of the $l$-th path for the $n$-th antenna. Then the time-invariant baseband channel impulse response of the $n$-th received antenna can be given by\cite{tse}
	\begin{equation}
		h_{n}(\tau) = \sum_{l=1}^{L}\tilde{\alpha}_l \kappa_{l,n}\mathrm{e}^{-\mathrm{j}2\pi f_c \tau_{l,n}}\delta(\tau-\tau_{l,n}),
		\label{h_ln}
	\end{equation}
	where $\tilde{\alpha}_l \in \mathbb{C}$ denotes the complex path gain. $\kappa_{l,n}$ is introduced to characterize the SnS properties from the perspective of multipath propagation mechanisms, and it is given by \cite{SnS1}
	\begin{equation}
		\kappa_{l,n}  
		\begin{cases}
			=0, \, n \notin \boldsymbol{\phi}_l,\\
			=1, \, n \in    \boldsymbol{\phi}_l \, \& \, \mathrm{LoS/Reflection},\\
			>0, \, n \in \boldsymbol{\phi}_l \, \& \, \mathrm{Diffraction},
		\end{cases}
		\label{SnS}
	\end{equation} 
	where $\boldsymbol{\phi}_l = [n_{l,s}, n_{l,e}]$ denotes the VR of the $l$-th path with $n_{l,s}$ and $n_{l,e}$ indicating the start and end index\footnote{For the diffraction paths, $\kappa_{l,n} \in \boldsymbol{\phi}_l$ is associated with the diffraction coefficient. As such, $\kappa_{l,n}$ can be set within $0 < \kappa_{l,n} < 1$, with the reference value chosen as the element with the highest received power. Thus, we assume $\kappa_{l,n} \in \boldsymbol{\phi}_l$ obeys a uniform distribution between 0 and 1. Additionally,  we assume continuous VRs for simplicity. Notably, the proposed method is not restricted to continuous VRs and can be extended to scenarios involving disjoint VRs, where the visible antennas may be distributed across multiple disjoint regions for a specific path.}. 
	Assume the distance and direction of the $l$-th path to the reference antenna element as $r_l$ and $\vartheta_l$, as illustrated  in Fig.~\ref{SystemModel}.
	Then, $\tau_{l,n}$, for any $l\in \mathcal{L}$, can be rewritten as
	\begin{equation}
		\tau_{l,n} = \frac{r_{l,n}}{c} = \tau_{l} - \frac{r_l-r_{l,n}}{c},
		\label{tau_ln}
	\end{equation}
	where $\tau_{l} = r_l/c$; $r_{l,n}$ denotes the distance between the $l$-th scatterer and the $n$-th antenna element, which is defined as
	\begin{equation}
		 \begin{aligned}
		 	r_{l, n} =& \sqrt{(r_l\cos\vartheta_l-nd)^2+r_l^2\sin^2\vartheta_l}\\
		 	\overset{(a)}{\approx}& r_l- nd \cos \vartheta_l + \frac{n^2d^2}{2r_l} {\sin^2\vartheta_l},
		 \end{aligned}
		\label{Delta_ln}
	\end{equation}
	where $(a)$ is obtained by applying the second-order Taylor approximation with the current point set at $n = 0$ and terms of third order and higher neglected. Notably, this approximation maintains high accuracy when the distance between the BS and the user or scatterer exceeds the Fraunhofer distance \cite{PolarCS}. Additionally, unlike the Euclidean-domain model in \cite{Rewiewer4}, the approximation enables the array response vector to align with a spatial chirp form, simplifying subsequent analysis.
	
	Utilizing (\ref{Delta_ln}), (\ref{tau_ln}) can be rewritten as $\tau_{l,n} = \tau_{l} - n{\psi_l}/{f_c} + n^2{\varphi_l}/{f_c}$, where
	$\psi_l \triangleq d\cos \vartheta_l / \lambda_c$ and $\varphi_l \triangleq {d^2\sin^2\vartheta_l}/{2r_l\lambda_c}$.
	According to (\ref{tau_ln}) and (\ref{Delta_ln}), the channel impulse response in (\ref{h_ln}) can be further expressed as
	\begin{equation}
		h_{n}(\tau) = \sum_{l=1}^{L}\alpha_l \kappa_{l,n}
		\mathrm{e}^{\mathrm{j}2\pi \left(n{\psi_l} - n^2{\varphi_l} \right)} \delta\left(\tau-\tau_{l,n}\right),
		\label{h_ln2}
	\end{equation}
	where $\alpha_l \triangleq \tilde{\alpha}_l\mathrm{e}^{-\mathrm{j}2\pi f_c \tau_{l}}$ is the equivalent complex path gain. 
	Applying the Fourier transform to (\ref{h_ln2}),
	the spatial-frequency response of the $n$-th antenna is given by (\ref{h_lnf})\cite{tse2,tse3}, as shown in the top of the next page.
	\begin{figure*}[!t]
		\normalsize
		\setcounter{MYtempeqncnt}{\value{equation}}
		\setcounter{equation}{8}
			\begin{equation}
				\begin{aligned}
					h_{n}(f) = \int_{-\infty}^{+\infty}h_{n}(\tau) \mathrm{e}^{-\mathrm{j}2\pi f\tau}\mathrm{d}\tau = \sum_{l=1}^{L}\alpha_l \kappa_{l,n}
					\mathrm{e}^{\mathrm{j}2\pi \left(n{\psi_l} - n^2{\varphi_l} \right)} \mathrm{e}^{\mathrm{j}2\pi f\left(n\frac{\psi_l}{f_c}- n^2\frac{\varphi_l}{f_c}\right)}\mathrm{e}^{-\mathrm{j}2\pi f\tau_{l}}.
				\end{aligned}
				\label{h_lnf}
			\end{equation}
		\hrulefill
		\vspace*{4pt}
	\end{figure*}
	Further, (\ref{h_lnf}) can be expressed in a more compact form as
	\begin{equation}
		\mathbf{h}(f) = \sum_{l=1}^{L}\alpha_l a(f, \tau_l)(\boldsymbol{\kappa}_l \odot \mathbf{b}(\psi_l, \varphi_l)) \odot \boldsymbol{\theta}(f, \psi_l, \varphi_l),
		\label{hvec}
	\end{equation}
	where $a(f, \tau_l) = \mathrm{e}^{-\mathrm{j}2\pi f\tau_{l}}$; $\boldsymbol{\kappa}_l = [\kappa_{l,1}, \kappa_{l,2}, \cdots, \kappa_{l,N}]^{\mathrm{T}}$ denotes the visibility indicator vector with the $n$-th entry being $\kappa_{l,n}$;
	$\mathbf{b}(\psi_l, \varphi_l)$ and $\boldsymbol{\theta}(f, \psi_l, \varphi_l)$ denote the array response vector and the frequency-dependent phase-shift vector, respectively, with each of their $n$-th entries given by
	\begin{align}
		\label{AAR1} [\mathbf{b}(\psi_l, \varphi_l)]_n &= \mathrm{e}^{\mathrm{j}2\pi \left(n{\psi_l} - n^2{\varphi_l} \right)},\\
		\label{AAR2} [\boldsymbol{\theta}(f, \psi_l, \varphi_l)]_n &= \mathrm{e}^{\mathrm{j}2\pi f\left(n\frac{\psi_l}{f_c}- n^2\frac{\varphi_l}{f_c}\right)}.
	\end{align}
	
	Assuming the total bandwidth of the OFDM system is denoted by $f_s$, the center frequency for each pilot subcarrier is given by $f_k = f_c + {kf_s}/{K}$ with $k = 0,1,\cdots,K-1$. The corresponding channel vector for each subcarrier is then represented as $\mathbf{h}(f_k)$, as illustrated in (\ref{hvec}) with $f$ replaced by $f_k$. Thus, the overall spatial-frequency channel matrix $\mathbf{H}$ can be reformulated as
	\begin{equation}
		\mathbf{H} = \sum_{l=1}^{L}\alpha_l
		\left(\boldsymbol{\kappa}_l\odot\mathbf{b}(\psi_l, \varphi_l)\right) \mathbf{a}^{\mathrm{T}}(\tau_l)\odot\boldsymbol{\Theta}(\psi_l, \varphi_l).
		\label{H_matrix}
	\end{equation}
	where $\mathbf{a}(\tau_{l}) = [a(f_1, \tau_l), \cdots, a(f_K, \tau_l)]^{\mathrm{T}}$, and $\boldsymbol{\Theta}(\psi_l, \varphi_l) = [\boldsymbol{\theta}(f_0, \psi_l, \varphi_l), \cdots, \boldsymbol{\theta}(f_{K-1}, \psi_l, \varphi_l)]$.
	
	\begin{remark}
		Eq. (\ref{H_matrix}) presents an approximated channel model for XL-MIMO systems, integrating spherical wavefront effects, SnS properties, as well as spatial and frequency-wideband characteristics. Specifically, the spatial-domain steering vector $\mathbf{b}(\psi_l, \varphi_l)$ is coupled with the SnS indicator vector $\boldsymbol{\kappa}_l$, the frequency response vector $\mathbf{a}(\tau_l)$, and the frequency-dependent phase matrix $\boldsymbol{\Theta}(\psi_l, \varphi_l)$. 
	\end{remark}
	\vspace{-1.5em}
	\subsection{Angular-Delay Representation of SnS XL-MIMO Channel}
	To enable efficient channel estimation, leveraging channel sparsity is critical. Regarding the spatial domain, the Euclidean and polar domains are theoretically preferable transformation choices where each propagation path in these domains may correspond to a single line spectrum, yielding a highly sparse representation\cite{TLGAMP}. However, these domains rely on joint two-dimensional sampling, which leads to transformation matrices with drastically high dimensionality, and consequently, heightened computational complexity.
		
	As an alternative, we propose leveraging angular-delay sparsity for estimation \cite{Dual_Wideband}. Thus, the spatial-frequency channels in (\ref{H_matrix}) can be approximated as
	\begin{equation}
		\mathbf{H} = \mathbf{F}_{\mathrm{A}}\mathbf{X}\mathbf{F}_{\mathrm{D}}^{\mathrm{T}},
		\label{H_DFT}
	\end{equation}
	 where $\mathbf{F}_{\mathrm{A}} \in \mathbb{C}^{N \times N}$ and $\mathbf{F}_{\mathrm{D}} \in \mathbb{C}^{K \times K}$ denote the normalized $N$- and $K$-dimensions discrete Fourier transformation (DFT) matrices. $\mathbf{X} \in \mathbb{C}^{N \times K}$ denotes the angular-delay channel.
	
	From (\ref{AAR1}) and (\ref{AAR2}), it can be observed that both the frequency-dependent and frequency-independent array response vectors take the form of spatial chirp signals. Consequently, they can be reformulated as
	\begin{align}
		\label{AAR3} [\mathbf{b}(\psi_l, \varphi_l)]_n &= \mathrm{e}^{\mathrm{j}2\pi \left({\psi_l} - n{\varphi_l} \right)n},\\
		\label{AAR4} [\boldsymbol{\theta}(f_k, \psi_l, \varphi_l)]_n &= \mathrm{e}^{\mathrm{j}2\pi \left(a_{l,k}n- \frac{b_{l,k}}{2}n\right)n},
	\end{align}
	where $a_{l,k} = \psi_l{f_k}/{f_c}$ and $b_{l,k} = 2\varphi_l{f_k}/{f_c}$ denote the initial spatial frequency and the spatial chirp rate, respectively.
	
	By exploiting the chirp-like properties described in (\ref{AAR3}), the start and end spatial frequencies of the SnS array response vector $\boldsymbol{\kappa}_l \odot \mathbf{b}(\psi_l, \varphi_l)$ can be approximated as \cite{Chirp1}
	\begin{equation}
		\mathcal{I}_{l,s} = \psi_l- 2\varphi_l {n_{l,s}}, \quad \mathcal{I}_{l,e} = \psi_l - 2\varphi_l {n_{l,e}}.   
		\label{Chirp1}
	\end{equation}
	In this manner, the array response vector $\boldsymbol{\kappa}_l \odot \mathbf{b}(\psi_l, \varphi_l)$, which captures both the spherical wavefront effect and the SnS characteristics, can be approximated as a superposition of multiple far-field array response components, i.e.,
	\begin{equation}
		\boldsymbol{\kappa}_l \odot \mathbf{b}(\psi_l, \varphi_l) = \int_{\mathcal{I}_{l,e}}^{\mathcal{I}_{l,s}} c(\epsilon_l) \mathbf{b}_{\mathrm{far}}(\epsilon_l) \mathrm{d}\epsilon_l,
		\label{dcom}
	\end{equation}
	where $c(\epsilon_l)$ denotes the decomposition coefficient, and $[\mathbf{b}_{\mathrm{far}}(\epsilon_l)]_n = \mathrm{e}^{\mathrm{j}2\pi \epsilon_l n}$ represents the far-field array response corresponding to spatial frequency $\epsilon_l \in [\mathcal{I}_{l,e}, \mathcal{I}_{l,s}]$. Furthermore, the continuous decomposition in (\ref{dcom}) can be discretized with a spatial frequency resolution of $2/N$ as 
	\begin{equation}
		\boldsymbol{\kappa}_l \odot \mathbf{b}(\psi_l, \varphi_l) = \sum_{p=1}^{P} c_{l,p} \mathbf{b}_{\mathrm{far}}(\epsilon_{l,p}),
		\label{dcom2}
	\end{equation}
	where the number of sampling points is given by $P = \lceil ({n_{l,s}} - {n_{l,e}})\varphi_l N \rceil$, and the discretized spatial frequency is defined as $\epsilon_{l,p} = \min(\mathcal{I}_{l,s}, \mathcal{I}_{l,e}) + {2(p-1)}/{N}$ for $p = \{1, 2, \dots, P\}$. The coefficient $c_{l,p}$ denotes the corresponding weight associated with the $p$-th far-field array response component.
	
	By utilizing the decomposition in (\ref{dcom2}), the spatial-frequency channel described in (\ref{H_matrix}) can be expressed as
	\begin{equation}
		\mathbf{H} = \sum_{l=1}^{L}\sum_{p=1}^{P}\alpha_lc_{l,p}
		\mathbf{b}_{\mathrm{far}}(\epsilon_{l,p}) \mathbf{a}^{\mathrm{T}}(\tau_l)\odot\boldsymbol{\Theta}(\psi_l, \varphi_l).
		\label{H_matrix2}
	\end{equation}
	
	Similar to the analysis in (\ref{Chirp1}), the start and end spatial frequencies of the frequency-dependent phase-shift vector $\boldsymbol{\theta}(f_k, \psi_l, \varphi_l)$ can be approximated as
	\begin{align}
		\label{chirp2} i_{l,s}(k) &= (a_{l,k}-b_{l,k}n_{l,s}){\lambda_\mathrm{c}}/{d},\\
		i_{l,e}(k) &= {(a_{l,k}-b_{l,k}n_{l,e})}\lambda_\mathrm{c} /{d}.
		\label{chirp3}
	\end{align}
	In addition, for notation simplification, we define 
	\begin{equation}
		t_{l,n} = \frac{n^2{f_s\varphi_l}-n{f_s\psi_l}}{Kf_c}.
	\end{equation}
	
	\textbf{Lemma 1.} By utilizing (\ref{H_matrix2})-(\ref{chirp3}), it can be shown that the two-dimensional inverse discrete Fourier transform (2D-IDFT) of $\boldsymbol{\Theta}(\psi_l, \varphi_l)$ results in a sparse matrix, where the nonzero entries are confined within a square region. Within this region, the dominant components exhibit a clear column-wise clustered sparsity structure. Specifically, as $N, K \rightarrow \infty$, the support of the nonzero region converges to:
	\begin{equation}
		 [\mathbf{F}_{\mathrm{A}}^{\mathrm{H}}\boldsymbol{\Theta}(\varphi_l,\psi_l)\mathbf{F}_{\mathrm{D}}^{\mathrm{*}}]_{n,k} = \begin{cases}
			\text{nonzeros}, &(n,k) \in \mathcal{A}_{l},\\
			0, &(n,k) \notin \mathcal{A}_{l},
		\end{cases} 
		\label{A1}
	\end{equation}
	where the set $\mathcal{A}_{l}$ is defined as 
	\begin{equation}
		\mathcal{A}_{l} \triangleq \left\{(n,k) \in \mathbb{Z}^{2} \mid I_{l,e} \le n \le I_{l,s}, J_{l,s} \le k \le J_{l,e} \right\},
		\label{region1}
	\end{equation}
	with
	\begin{equation}
		\begin{aligned}
			I_{l,e} &= \min_k \lceil (i_{l,e}(k)+1)N/2 \rceil, \\
			I_{l,s} &= \max_k \lceil (i_{l,s}(k)+1)N/2 \rceil, \\
			J_{l,s} &= \min_n j_{l,s}(n), \, J_{l,e} = \max_n j_{l,e}(n),
		\end{aligned}
	\end{equation}
	indicating the boundaries with $j_{l,s}(n) = (t_n - {2}/{K}+1/2)K$ and $j_{l,e}(n) =(t_n + {2}/{K}+1/2)K$.
	\begin{proof}
		Please see Appendix A.
	\end{proof}
	
	Within the region $\mathcal{A}_l$, each column corresponds to the angular response associated with a specific propagation delay. The width of the significant spatial frequency components is primarily determined by the frequency-width effect. Assuming the starting spatial frequency of the $k$-th column is denoted by $i_{l,k}\in [\min_k i_{l,e}(k), \max_k i_{l,s}(k)]$, the corresponding angular spread can be approximated as
	\begin{equation}
		B_{l,k} = \frac{i_{l,k}f_s}{f_c}N.
		\label{Bs}
	\end{equation}
	As a result, the significant components of $\mathbf{F}_{\mathrm{A}}^{\mathrm{H}}\boldsymbol{\Theta}(\varphi_l,\psi_l)\mathbf{F}_{\mathrm{D}}^{\mathrm{*}}$ within $\mathcal{A}_l$ are confined to narrow bands along each column, thereby exhibiting a pronounced column-wise clustered sparsity pattern.
	To summarize, by leveraging (\ref{A1})-(\ref{Bs}), the sparse structure of $\mathbf{F}_{\mathrm{A}}^{\mathrm{H}}\boldsymbol{\Theta}(\varphi_l,\psi_l)\mathbf{F}_{\mathrm{D}}^{*}$ can be accurately characterized. In particular, (\ref{A1}) determines the overall boundaries of the nonzero region, while (\ref{Bs}) provides a more precise description of the angular spread in each column, further refining the support of the dominant components.

	\textbf{Theorem 1.} Based on \textbf{Lemma 1}, the angular-delay domain channel matrix $\mathbf{X} = \mathbf{F}_{\mathrm{A}}^{\mathrm{H}}\mathbf{H}\mathbf{F}_{\mathrm{D}}^{*}$ exhibits a sparse structure, containing only $L$ nonzero blocks, each corresponding to an individual propagation path. For the $l$-th path, the support region of the nonzero components is denoted by $\mathcal{B}_l$, which can be expressed as $\mathcal{B}_{l} = \bigcup_{p} \mathcal{B}_{l,p}$, where $\mathcal{B}_{l,p}$ represents the shifted version of $\mathcal{A}_l$ in both the angular and delay domains, and is defined as
	\begin{equation}
		\begin{aligned}
			\mathcal{B}_{l,p} &= \left\{(n,k)\in\mathbb{Z}^2 \mid n = \mathrm{mod}(n_{l} + N\epsilon_{l,p}, N), \right.\\ 
			k &\left.= \mathrm{mod}(k_l + f_s\tau_l, K), \forall (n_l, k_l) \in \mathcal{A}_{l}\right\},
		\end{aligned}
		\label{region2}
	\end{equation}
	where $\mathrm{mod}(a, m)$ is the modulus of $a$ for $m$.
	\begin{proof}
		Please see Appendix B.
	\end{proof}

	According to \textbf{Theorem 1}, the nonzero region of $\mathbf{X}$ is formed by aggregating the shifted nonzero regions of $\mathbf{F}_{\mathrm{A}}^{\mathrm{H}}\boldsymbol{\Theta}(\psi_l, \varphi_l)\mathbf{F}_{\mathrm{D}}^*$ across both the angular and delay domains for all paths $l$. The shift positions are jointly determined by the spatial shift $N\epsilon_{l,p}$ and the delay shift $f_s \tau_l$, corresponding to the spatial frequency and propagation delay of each path, respectively.
	
	Moreover, we analyze the effects of the spherical wavefront, SnS properties, and dual-wideband effects on the sparsity structure of $\mathbf{X}$ based on \textbf{Lemma 1} and \textbf{Theorem 1} from the perspectives of the angular and delay spread. 
	\begin{itemize}
		\item \textbf{Angular Spread}: First, the spherical wavefront effect and the SnS characteristics determine the angular spread range at each subcarrier, characterized by $i_{l,s}(k)$ and $i_{l,e}(k)$. Then, the frequency-wideband effect governs the overall boundaries of the angular spread, denoted by $I_{l,e}$ and $I_{l,s}$, and enables a more accurate description of the angular spread within each column, represented by $B_{l,k}$.
		\item \textbf{Delay Spread}: The spatial-wideband effect causes variations in $j_{l,e}(n)$ and $j_{l,s}(n)$ across antennas $n$, which in turn leads to delay spread. The SnS properties define the range of variation for the antenna index $n$, while the spherical wavefront imposes a quadratic trend on the delay spread as a function of $n$.
	\end{itemize}

	To summarize, the spherical wavefront effect, SnS properties, and dual-wideband effects collectively govern the angular and delay spreads, thereby reshaping the sparsity structure of XL-MIMO channels in the angular-delay domain. Specifically, 
	\begin{itemize}
		\item \textbf{Global Block Sparsity:} The angular-delay domain channel exhibits $L$ nonzero blocks, each corresponding to a distinct propagation path. The index range of each nonzero block is characterized by $\mathcal{B}_{l}$.
		\item \textbf{Local Common-Delay Sparsity:} Due to the frequency-wideband effect, the significant components within $\mathcal{B}_l$ are confined to narrow bands along each column, thereby exhibiting pronounced column-wise clustered structures with strong correlation.
	\end{itemize}

	\begin{remark}
		The global block sparsity and local common-delay sparsity arise from the combined effects of the spherical wavefront, SnS properties, and dual-wideband characteristics. These features inherently embed inter-antenna and inter-subcarrier correlations into the angular-delay channel.
		Furthermore, (\ref{H_DFT}) offers an alternative representation of the spatial-frequency channel, facilitating the estimation of SnS dual-wideband XL-MIMO channels. Once the angular-delay channel is reconstructed, key parameters such as AoAs, VRs, and the range of beam squint can be efficiently extracted.
	\end{remark}
	\begin{figure*}
		\centering
		\subfigure[Angular spread for $k= 0$ and $K-1$]{
			\includegraphics[width=0.31\textwidth]{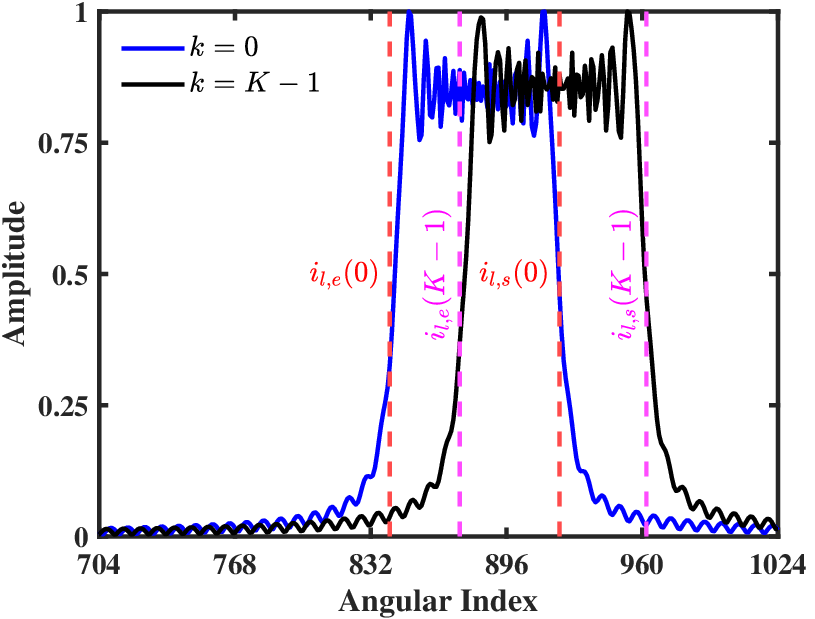}
			\label{LMF16}
		}
		\subfigure[Delay spread for $n= n_s$ and $n_e$]{
			\includegraphics[width=0.3\textwidth]{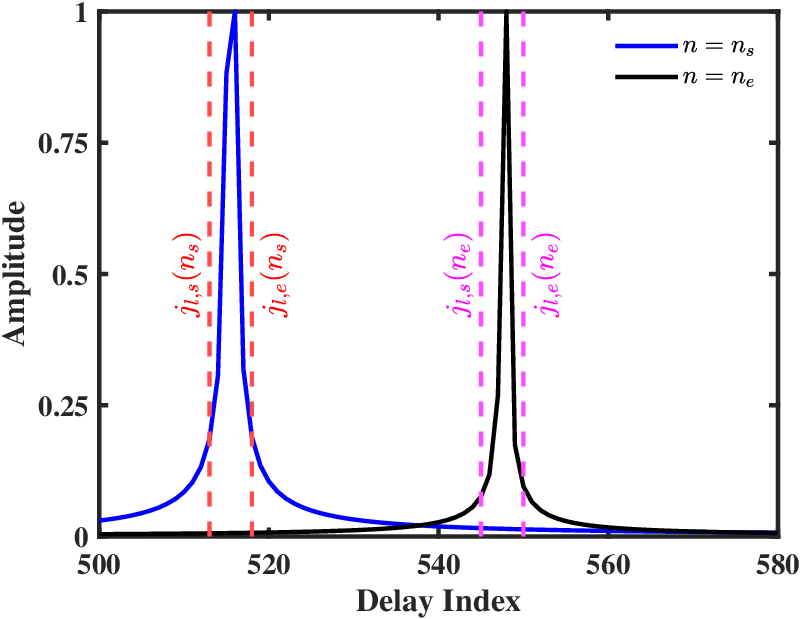}
			\label{LMF17}
		}
		\quad
		\subfigure[$\mathcal{A}_{l}$.]{
			\includegraphics[width=0.3\textwidth]{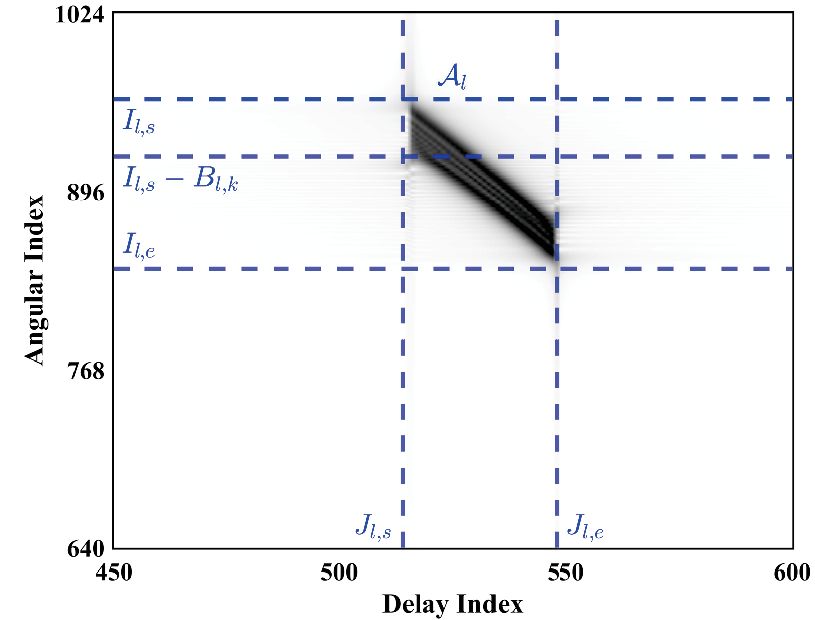}
			\label{LMF12}
		}

		\subfigure[$\mathcal{B}_{l,1}$.]{
			\includegraphics[width=0.3\textwidth]{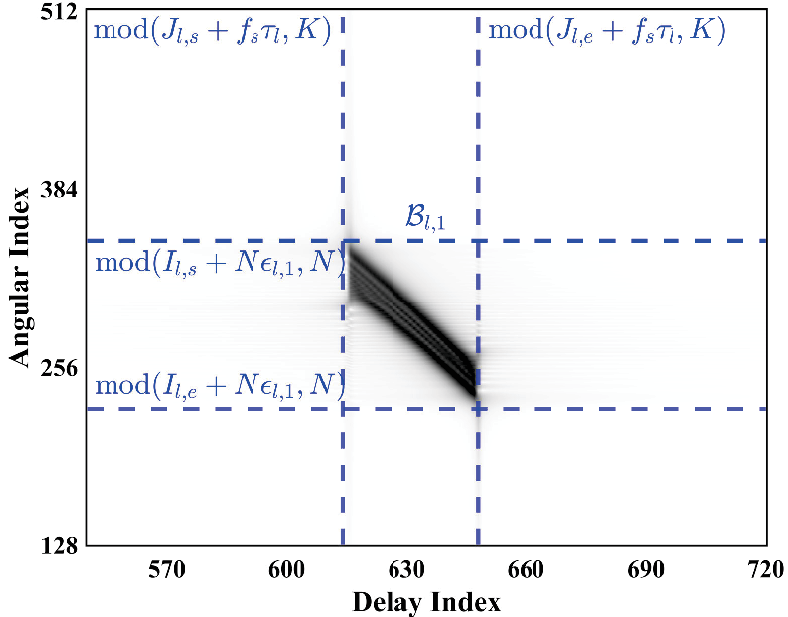}
			\label{LMF13}
		}
		\quad
		\subfigure[$\mathcal{B}_{l,P}$]{
			\includegraphics[width=0.3\textwidth]{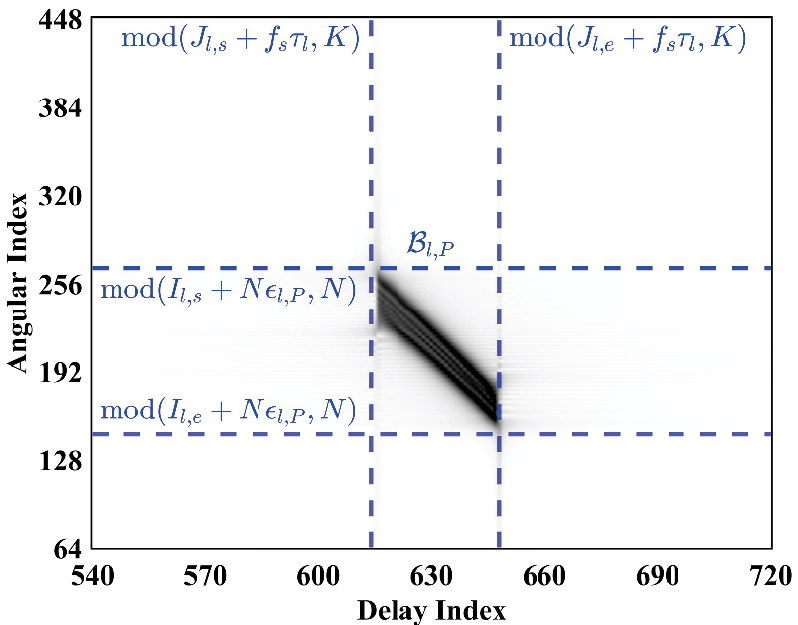}
			\label{LMF14}
		}
		\quad
		\subfigure[$\mathcal{B}_{l}$.]{
			\includegraphics[width=0.3\textwidth]{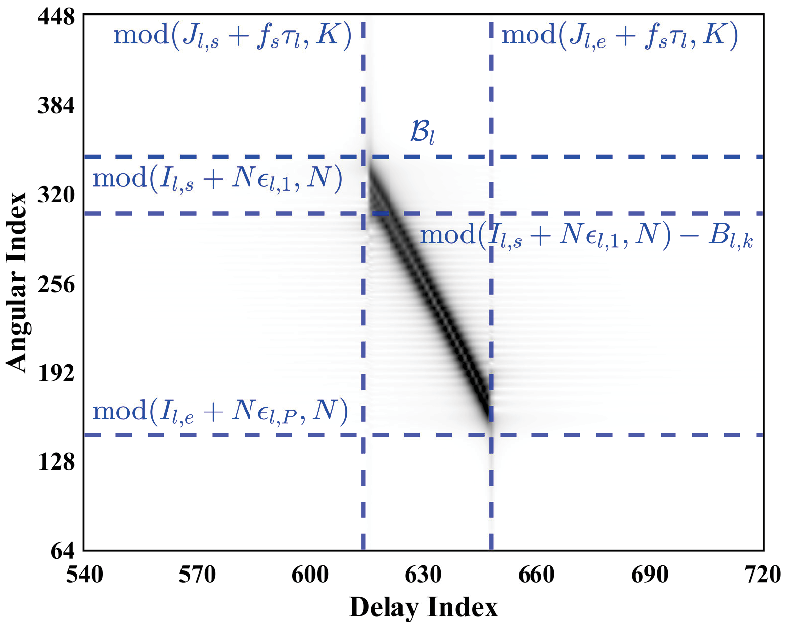}
			\label{LMF15}
		}
		\caption{Illustration of block sparsity of SnS XL-MIMO channel in the angular-delay domain.}
		\label{spectrum} 
	\end{figure*}
	
	{To enhance intuitive understanding of the XL-MIMO channel, we provide an example with the following parameters: $N=K=1024$, $f_s = 3$GHz, $f_c = 30$GHz, $r_l = 10$m, $\vartheta_l = \pi/5$, and $\boldsymbol{\phi}_l \in [64,959]$. 
	Fig. \ref{spectrum} illustrates the sparsity pattern of the SnS XL-MIMO channel in the angular-delay domain.
	Given the specified parameters, we have $i_{l,e}(0) = 0.64$, $i_{l,s}(0) \approx 0.8$, $j_{l,s}(n_s) = 513$, and $j_{l,e}(n_s) \approx 517$. Similarly, $i_{l,e}(K-1) \approx 0.72$, $i_{l,s}(K-1) \approx 0.88$,  $j_{l,s}(n_e) = 545$, and $j_{l,e}(n_e) \approx 549$. As depicted in Fig. \ref{LMF16} and Fig. \ref{LMF17}, these values provide accurate boundaries for the ranges of angular and delay spread.}
	
	Furthermore, we obtain the following values:
	$I_{l,e} = 841$,
	$I_{l,s} = 962$,
	$J_{l,s} = 513$,
	$J_{l,e} = 549$, and 
	$B_{k,s} \approx 41$ with $k = J_{l,e}$.
	Fig. \ref{LMF12} illustrates the non-zero region of $\mathbf{F}_{\mathrm{A}}^{_{\mathrm{H}}}\boldsymbol{\Theta}(\psi_l, \varphi_l)\mathbf{F}^{*}_{\mathrm{D}}$, which is accurately constrained to $\mathcal{A}_{l}$, validating the effectiveness of \textbf{Lemma 1}. In particular, for the $J_{l,s}$-th column, the width of angular spread also aligns the result of (\ref{Bs}), which further validates the influence of frequency-wideband effect for local angular spread.
	
	Additionally, we have $f_s\tau_l = 100$, $\epsilon_{l,1} = \psi_l - 2n_{l,s}\varphi_l \approx 0.4$, and $\epsilon_{l,P} = \psi_l - 2n_{l,e}\varphi_l = 0.32$, corresponding to $N\epsilon_{l,1} = 410$ and $N\epsilon_{l,P} = 328$, respectively.
	Consequently, Fig. \ref{LMF13} and Fig. \ref{LMF14} represent shifted versions of Fig. \ref{LMF12}. The shift distances in the angular domain are $696$ and $614$, respectively, while the shift distance in the delay domain is $100$. 
	Finally, by concentrating the shifted versions across all $\epsilon_{l,p}$, the sparse structure depicted in Fig. \ref{LMF15} is obtained.

	\begin{remark}
		Compared to massive MIMO-OFDM channels, the wideband XL-MIMO channel exhibits distinct behavior. Specifically, in addition to beam squint, the spherical wavefront effect and SnS properties contribute to angular spread, resulting in broader angular spread. Moreover, the quadratic phase variations introduced by the spherical wavefront reshape the trend of delay variations across antennas.
		{Notably, \textbf{Lemma 1} and \textbf{Theorem 1}  are applicable to massive MIMO-OFDM channels. When the spherical wavefront simplifies to a plane wavefront and SnS properties are disregarded, both $\varphi_l$ and $\kappa_{l,n}$ can be omitted. Thus, we have $i_{l,e}(k) = i_{l,s}(k)$ and $t_{l,n} = n f_s \psi_{l}$. Consequently, the significant components are uniformly distributed in the non-zero region \cite{Dual_Wideband}.}
	\end{remark}
	\section{Problem Formulation and Sparse Prior Modeling}
	\label{section4}
	In this section, leveraging the sparsity outlined in \textbf{Theorem 1}, we first formulate the XL-MIMO channel estimation problem as a sparse recovery task. Subsequently, to capture the global block sparsity and local common-delay sparsity, we introduce a column-wise sparse prior model.
	\vspace{-1em}
	\subsection{Problem Formulation}
	Motivated by the sparsity of XL-MIMO channels in the angular-delay domain, the channel estimation problem can be formulated as a MMV-based sparse recovery problem. Utilizing the angular-delay representation in (\ref{H_DFT}), the received signal model in (\ref{Y_pilot}) can be further written as
	\begin{equation}
		\mathbf{Y} = \mathbf{W}\mathbf{F}_{\mathrm{A}}\mathbf{X}\mathbf{F}_{\mathrm{D}}^{\mathrm{T}} + {\mathbf{N}} = \boldsymbol{\Psi}\mathbf{X}\mathbf{F}_{\mathrm{D}}^{\mathrm{T}} + {\mathbf{N}},
		\label{Y_pilot2}
	\end{equation}
	where $\boldsymbol{\Psi}\triangleq \mathbf{W}\mathbf{F}_{\mathrm{A}} \in \mathbb{C}^{M\times N}$. Furthermore, utilizing the unitary property of $\mathbf{F}_{\mathrm{D}}$, (\ref{Y_pilot2}) can be reformulated as
	\begin{equation}
		\tilde{\mathbf{Y}} = {\mathbf{Y}} \mathbf{F}_{\mathrm{D}}^{*}= \boldsymbol{\Psi}\mathbf{X} + \tilde{\mathbf{N}},
		\label{Y_pilot3}
	\end{equation}
	where $\tilde{\mathbf{Y}} = {\mathbf{Y}} \mathbf{F}_{\mathrm{D}}^{*} \in \mathbb{C}^{M \times K}$ and $\tilde{\mathbf{N}} = {\mathbf{N}}\mathbf{F}_{\mathrm{D}}^{*} \in \mathbb{C}^{M \times K}$ denotes the equivalent received pilot signal and  noise matrix.

	\begin{remark}
		While this work primarily focuses on the ULA configuration, the signal model in  (\ref{Y_pilot3}) is also applicable to the uniform planar array (UPA) configuration. Specifically, consider a half-wavelength UPA comprising $N = N_v \times N_h$ antennas, where $N_v$ and $N_h$ denote the number of elements along the vertical and horizontal dimensions, respectively. In this case, the angular transformation matrix $\mathbf{F}_{\mathrm{A}}$ can be redefined as $\mathbf{F}_{\mathrm{A}} = \mathbf{F}_{\mathrm{A},v} \otimes \mathbf{F}_{\mathrm{A},h}$, where $\mathbf{F}_{\mathrm{A},v} \in \mathbb{C}^{N_v \times N_v}$ and $\mathbf{F}_{\mathrm{A},h} \in \mathbb{C}^{N_h \times N_h}$ denote the vertical and horizontal DFT matrices, respectively.
	\end{remark}
	\vspace{-0.5em}
	This paper aims to develop an effective estimation algorithm for reconstructing $\mathbf{H}$ according to $\mathbf{Y}$ and $\boldsymbol{\Psi}$.
	So far, several sparse signal recovery algorithms have been employed to address the problem in (\ref{Y_pilot3}), including SOMP \cite{PolarCS, OMP_1} and optimization-based methods \cite{Optimization_1, atomic}. However, these methods fail to effectively exploit the inherent sparsity structure of the angular-delay channel and often require additional knowledge, such as the rank of $\mathbf{X}$ or the number of multipaths, which may not always be available.
	Considering the block sparsity of $\mathbf{X}$, Bayesian inference techniques such as approximate message passing (AMP) and SBL have shown superior recovery performance in channel estimation tasks \cite{GAMP3, GAMP5, Variance_State_2}. However, the effectiveness of these Bayesian methods relies heavily on the accuracy of the prior model.
	\vspace{-1em}
	\subsection{Column-Wise Hierarchical Sparse Prior Model}
	According to \textbf{Lemma 1} and \textbf{Theorem 1}, the angular-delay domain channel exhibits prominent block sparsity and a column-wise clustered sparsity structure. However, existing sparse prior models fail to capture these characteristics fully \cite{GAMP5, GAMP3, Variance_State_1, Variance_State_2}. 
	For example, the two-layer hierarchical prior models in \cite{GAMP5,GAMP3} assign independent priors to the precision parameters of the channel coefficients, which limits their ability to model the correlation across coefficients.
	Although \cite{Variance_State_1,Variance_State_2} leverage MRF to promote global block sparsity, they still overlook local structural dependencies, especially those induced by column-wise clustering. As a result, these models are not well suited for the XL-MIMO channel estimation problem considered in this work.
	
	To this end, we propose a novel column-wise hierarchical sparse prior model, which incorporates both precision sharing mechanism and MRF structure to promote structured sparsity. Specifically, the column-wise hierarchical prior is defined as
	\begin{equation}
		p(\mathbf{X}, \boldsymbol{\Gamma}, \boldsymbol{\Omega}, \mathbf{S}) = p(\mathbf{X}| \boldsymbol{\Gamma}) p( \boldsymbol{\Gamma}| \boldsymbol{\Omega}, \mathbf{S})p(\boldsymbol{\Omega})p(\mathbf{S}).
		\label{prior1}
	\end{equation}
	We next detail the distributions associated with each layer.
	
	\textbf{1) Variance-Driven Sparsity}: In the first layer, we adopt a complex Gaussian prior with mean zero and variance $[\boldsymbol{\Gamma}]_{n,k} = \gamma_{n,k}^{-1}$ for each coefficient, i.e., 
	\begin{equation}
		\label{prior2} p(\mathbf{X}| \boldsymbol{\Gamma}) = \prod_{n=1}^{N}\prod_{k=1}^{K}\mathcal{CN}(x_{n,k}; 0, \gamma_{n,k}^{-1}),
	\end{equation}
	where $\gamma_{n,k}^{-1}$ denotes the variance corresponding to coefficient $x_{n,k}$. This variance-driven modeling naturally induces sparsity. Specifically, as the variance approaches zero, the corresponding $x_{n,k}$ is effectively pushed toward zero, thereby promoting a sparse solution.
	
	\textbf{2) Precision Sharing Mechanism}: 
	Considering the sparsity of angular-delay channel, we adopt a dual-precision strategy, where each $\gamma_{n,k}$ is associated with two distinct precision parameters corresponding to the active and inactive states of channel coefficients. Building on this, to capture the local common-delay clustered sparsity, we further introduce the precision  sharing mechanism, where $\gamma_{n,k}$ is assigned a conditionally Bernoulli-Gamma distribution as
	\begin{align}
		\label{prior3} p(\boldsymbol{\Gamma}| \boldsymbol{\Omega}, \mathbf{S})&=\prod_{n=1}^{N}\prod_{k=1}^{K}\delta\left(\gamma_{n,k}-t(s_{n,k}, \boldsymbol{\alpha}_{k})\right),
	\end{align}
	where $t(s_{n,k}, \boldsymbol{\alpha}_k) = \delta(1-s_{n,k})\alpha_k^1 + \delta(1+s_{n,k})\alpha_k^2$ with $\boldsymbol{\alpha}_k = [\alpha_k^1, \alpha_k^2]^{\mathrm{T}}$ and $\delta(\cdot)$ indicating the Dirac delta function.
	The precision parameter $\boldsymbol{\Omega} = [\boldsymbol{\alpha}_1, \boldsymbol{\alpha}_2, \cdots, \boldsymbol{\alpha}_K]^{\mathrm{T}} \in \mathbb{R}^{2 \times K}$ of $\boldsymbol{\Gamma}$ is characterized by
	\begin{equation}
		\label{prior4} p(\boldsymbol{\Omega}) = \prod_{i=1}^{2}\prod_{k=1}^{K}p(\alpha_k^i)=\prod_{i=1}^{2}\prod_{k=1}^{K}\mathrm{Ga}(\alpha_k^i; a_i, b_i).
	\end{equation}
	As described in (\ref{prior3}) and (\ref{prior4}), this common-delay sparsity mechanism ensures that, for a fixed $k$, all $x_{n,k}$ sharing the same state are governed by a common precision parameter. This coupling enforces a consistent clustered sparsity structure among components associated with the same delay, which aligns with the local common-delay sparsity.

	\textbf{3) MRF-Based Variance State Modeling}: In the third layer, to capture the global block sparsity, the variance state variable $\mathbf{S}$ is modeled as a MRF
	\begin{equation}
		\label{prior5} p(\mathbf{S}) = \left(\prod_{n,k} \prod_{(n',k')\in \mathcal{D}_{n,k}}u(s_{n,k},s_{n',k'})\right)^{\frac{1}{2}}\prod_{n,k}v(s_{n,k}),
	\end{equation}
	where $\mathcal{D}_{n,k}$ denotes the set of neighboring nodes of $s_{n,k}$; $u(s_{n,k}, s_{n',k'}) = \exp(\varpi s_{n,k} s_{n',k'})$ and $v(s_{n,k}) = \exp(-\eta s_{n,k})$ represent the pairwise potential and self-potential functions, respectively, with $\varpi$ and $\eta$ being the model parameters associated with $p(\mathbf{S})$. By leveraging the formulation in (\ref{prior5}), the prior encourages block patterns in the support structure and suppresses isolated coefficients that deviate from their neighboring states.
	\vspace{-1em}
	\section{Proposed MMV-HMP Algorithm}
	\label{section5}
	Based on the proposed column-wise hierarchical sparse prior model, this section first formulates the sparse signal recovery problem as a MMV-based Bayesian inference task. Then, to effectively perform the inference, we propose an MMV-HMP algorithm.
	\vspace{-1em}
	\subsection{Bayesian Inference}
	Since the measurement matrix $\boldsymbol{\Psi}$ may be ``bad'' (e.g., rank-deficient, ill-conditioned, or having a non-zero mean) \cite{GAMP3}, the divergence issues might be arisen in the Bayesian inference. To address this, we first perform unitary transformations on the received signal. Let the singular value decomposition (SVD) of the measurement matrix be denoted as $\boldsymbol{\Psi} = \mathbf{U}\boldsymbol{\Lambda}\mathbf{V}^{\mathrm{H}}$, where $\mathbf{U}$ and $\mathbf{V}$ are two unitary matrices. Performing a unitary transformation with $\mathbf{U}^{\mathrm{H}}$  on the received signal model in (\ref{Y_pilot3}) yields the following model
	\begin{equation}
		\mathbf{R}  = \mathbf{A}\mathbf{X}+ {\mathbf{W}} =\mathbf{Z}+ {\boldsymbol{\Xi}},
		\label{Unitary_R}
	\end{equation}
	where $\mathbf{R} = \mathbf{U}^{\mathrm{H}}\tilde{\mathbf{Y}}$, $\mathbf{A}=\boldsymbol{\Lambda}\mathbf{V}^{\mathrm{H}}$, $\mathbf{Z} = \mathbf{A}\mathbf{X}$, and $\boldsymbol{\Xi} = \mathbf{U}^{\mathrm{H}}{\tilde{\mathbf{N}}}$.
	
	Based on the prior model provided in (\ref{prior1}), the maximum a posterior (MAP) estimator for the $(n,k)$-th entry of $\mathbf{X}$ can be expressed as
	\begin{equation}
		\hat{x}_{n,k}=\int {x}_{n,k} p(\boldsymbol{\Theta}|\mathbf{R}) \mathrm{d}\beta \mathrm{d}\mathbf{Z}\mathrm{d}\mathbf{X} \mathrm{d}\mathbf{\boldsymbol{\Gamma}} \mathrm{d}\mathbf{\boldsymbol{\Omega}}\mathrm{d}\mathbf{S},
		\label{Post_Estimation}
	\end{equation}
	where $\boldsymbol{\Theta}\triangleq\{\beta, \mathbf{Z}, \mathbf{X},\boldsymbol{\Gamma}, \boldsymbol{\Omega}, \mathbf{S}\}$, and 
	$p(\boldsymbol{\Theta}|\mathbf{R})$ denotes the joint posterior probability, which is defined as
	\begin{equation}
		\begin{aligned}
			p(\boldsymbol{\Theta}|\mathbf{R}) \propto  p(\mathbf{R}|\mathbf{Z}, \beta)p(\mathbf{Z}|\mathbf{X})p(\mathbf{X}, \boldsymbol{\Gamma}, \boldsymbol{\Omega}, \mathbf{S})p(\beta), 
		\end{aligned}
		\label{post}
	\end{equation}
	where $p(\beta) \propto \beta^{-1}$ denotes the prior distribution of noise precision; the conditional distributions $p(\mathbf{R}|\mathbf{Z}, \beta)$ and $p(\mathbf{Z}|\mathbf{X})$ are respectively given by
	\begin{align}
		p(\mathbf{R}|\mathbf{Z}, \beta) &= \prod_m\prod_kp(r_{m,k}\mid z_{m,k}),\\
		p(\mathbf{Z}|\mathbf{X}) &= \delta(\mathbf{Z}-\mathbf{AX}),
	\end{align}
	where $p(r_{m,k}|z_{m,k}) = \mathcal{CN}(z_{m,k};r_{m,k},\beta^{-1})$; ${r}_{m,k}$ and ${z}_{m,k}$ denote the $(m,k)$-th elements of $\mathbf{R}$ and $\mathbf{Z}$.
	
	Due to the large number of antennas and subcarriers in XL-MIMO systems, solving problem (\ref{Post_Estimation}) requires evaluating high-dimensional integrals, which is computationally prohibitive. Moreover, traditional SBL-based methods are infeasible in this context as they involve high-dimensional matrix inversions. Consequently, this highlights the pressing need for novel channel estimation techniques capable of addressing the challenges posed by the SnS dual-wideband channel estimation in XL-MIMO systems. Recently, message passing-based techniques have been widely adopted for solving MAP estimation problems, owing to their computational efficiency. Motivated by these advances, we propose a MMV-HMP algorithm to efficiently solve problem (\ref{Post_Estimation}).
	\vspace{-1em}
	\subsection{Factor Graph Representation} 
	\begin{table}[h]
		\centering
		\caption{Factor and Distribution in (\ref{post})}
		\setlength{\tabcolsep}{4mm}{
			\begin{tabular}{c c c}
				\toprule
				\makecell[c]{Factor} & \makecell[c]{Distribution} &  \makecell[c]{Function} \\
				\midrule
				$f_\beta$ &$p(\beta)$  & \makecell[c]{$\beta^{-1}$}  \\
				
				$f_{r_{m,k}}$& $p(r_{m,k}|z_{m,k},\beta)$  & \makecell[c]{$\mathcal{CN}(r_{m,k}; z_{m,k}, \beta^{-1})$} \\
				
				$f_{z_{m,k}}$& $p(z_{m,k}|\mathbf{x}_k)$  & \makecell[c]{$\delta(z_{m,k}-\mathbf{A}_{m,:}\mathbf{x}_k)$} \\
				
				$f_{x_{n,k}}$& $p(x_{n,k}|\gamma_{n,k})$  &$\mathcal{CN}(x_{n,k}; 0, \gamma_{n,k}^{-1})$   \\
				
				$f_{\gamma_{n,k}}$& $p(\gamma_{n,k}|\boldsymbol{\alpha}_k,s_{n,k})$ &$\delta(\gamma_{n,k}-t(s_{n,k}, \boldsymbol{\alpha}_k))$  \\
				\bottomrule
			\end{tabular}
		}
		\label{proba}
	\end{table}
	
\begin{figure*}[h]
		\centering
		\includegraphics[width=0.7\textwidth]{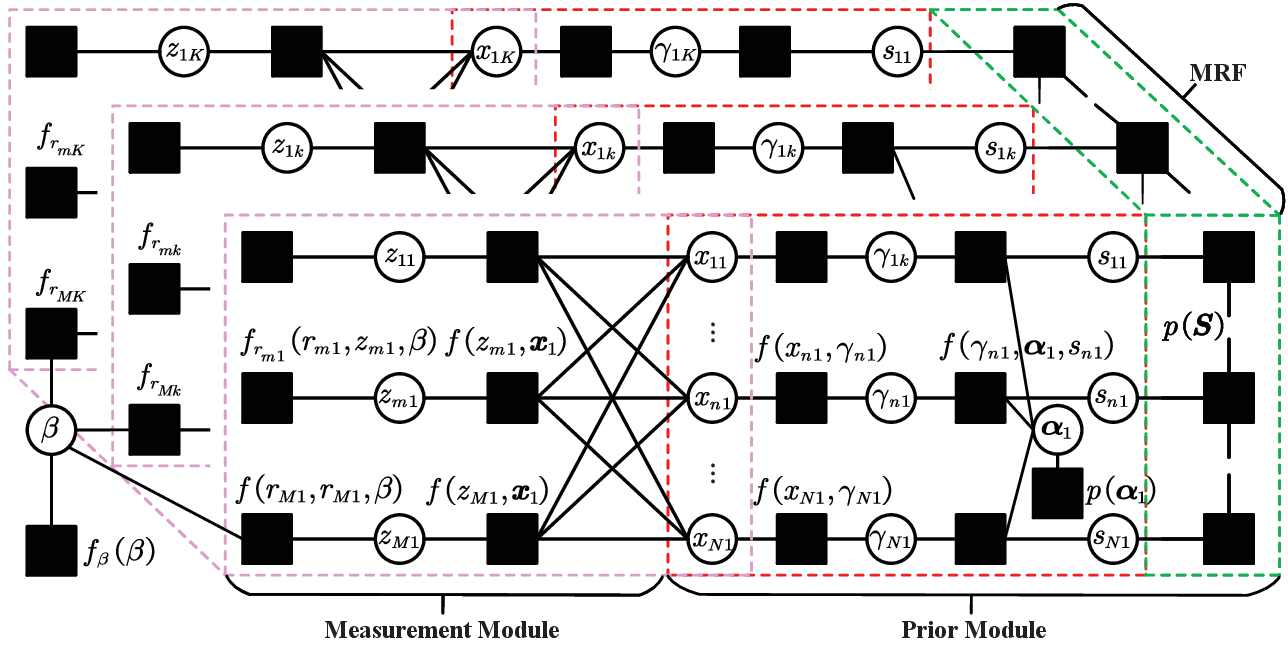}
		\caption{Factor graph representation for the factorization
			(\ref{post}).}
		\label{GF1}
	\end{figure*}

	The dependencies among the random variables in the factorization (\ref{post}) are illustrated in Fig. \ref{GF1}, with the corresponding probability distributions summarized in Table \ref{proba}. It is evident that the factor graph in Fig.~\ref{GF1} is significantly more intricate than those considered in \cite{GAMP3, GAMP4, GAMP5, Variance_State_1, Variance_State_2}, primarily due to the incorporation of precision parameter sharing mechanism and MRF structure. Specifically, the works in \cite{GAMP3, GAMP4, GAMP5} adopt independent priors on the variances of coefficients, thereby promoting element-wise sparsity. However, such models fail to capture the angular or delay-domain correlations among the coefficients. Furthermore, the model in \cite{Variance_State_1} does not include a precision parameter sharing mechanism, and thus overlooks higher-order dependencies among variable nodes. Due to these structural distinctions, existing message passing algorithms developed in \cite{GAMP3, GAMP4, GAMP5, Variance_State_1, Variance_State_2} are not directly applicable to our setting. Consequently, the message update equations at the variable nodes must be carefully reformulated to accommodate the enhanced complexity of our factor graph.
	
	Moreover, in the proposed sparse prior model, the latent precision parameters $\beta$ and $\alpha^{i}_k$ follow non-Gaussian distributions. 
	Meanwhile, the shared precision parameter simultaneously influences the message updates in both the first and third layers, resulting in inter-layer dependencies, which significantly increase the complexity of inference and render the exact computation of the node beliefs computationally intractable under a pure SP framework.
	
	To overcome these challenges, we propose a novel MMV-HMP algorithm jointly leveraging VMP and SP rules. Specifically, tailored to the intractable structure of the factor graph, we reformulate the message update equations along the edges, selectively applying SP or VMP rules to different edges to enable more flexible and efficient message updates.
	
	The MMV-HMP algorithm iteratively updates messages between adjacent nodes and aggregates them at the nodes $x_{n,k}$ to compute the corresponding posterior distributions, thereby circumventing the computationally intensive high-dimensional integrals in (\ref{Post_Estimation}). In the following sections, we introduce the details of the forward and backward message passing in measurement and prior modules.
	\vspace{-1em}
	\subsection{Measurement Module}
	Measurement module aims to obtain the likelihood estimation of $x_{n,k}$ based on the observation $\mathbf{R}$.
	Denote the belief of $z_{m,k}$ as $b(z_{m,k})\sim\mathcal{CN}(z_{m,k}; \hat{z}_{m,k}, \hat{\nu}^z_{m,k})$, which is defined in (\ref{bz_m}) with $\hat{z}_{m,k}$ and $\hat{\nu}^z_{m,k}$ indicating the $m$-th elements of $\hat{\mathbf{z}}_k$ and $ \hat{\boldsymbol{\nu}}_{\mathbf{z}_k}$, respectively. In this manner, according to the VMP rule \cite{VMP04,VMP05}, the message from $f_{r_{m,k}}$ to $\beta$ is given by
	\begin{align}
		\label{f_r2beta} m_{f_{r_{m,k}}\rightarrow \beta}(\beta) &\propto \exp\left\{\int \ln f_{r_{m,k}}b(z_{m,k})\mathrm{d}z_{m,k}\right\},\\
		&\overset{(a)}{\propto}\beta\exp\left\{-\beta\left(\left|r_{m,k}-\hat{z}_{m,k}\right|^2+\hat{\nu}^z_{m,k}\right)\right\},\notag
	\end{align}
	where $(a)$ is obtained by utilizing the first-order and second-order moment properties of Gaussian distribution.
	
	Concentrating all messages that are input to $\beta$, the belief of $\beta$ is given by
	\begin{align}
		\label{b_beta} &b(\beta) \propto f_{\beta}(\beta)\prod_{m=1}\prod_{k=1}m_{f_{r_{m,k}}\rightarrow \beta}(\beta),\\
		\notag \propto&\beta^{MK-1}\exp\left\{-\beta\sum_{m=1}\sum_{k=1}\left(\left|r_{m,k}-\hat{z}_{m,k}\right|^2+\hat{\nu}^z_{m,k}\right)\right\},
	\end{align}
	According to (\ref{b_beta}), it is observed that the belief $b(\beta)$ obeys the Gamma distribution with shape parameters $MK$ and $\sum_{m=1}\sum_{k=1}\left(\left|r_{m,k}-\hat{z}_{m,k}\right|^2+\hat{\nu}^z_{m,k}\right)$. Thus, the approximate posterior
	mean $\hat{\beta} = \int\beta b(\beta) \mathrm{d}\beta$ is given by
	\begin{equation}
		\hat{\beta} = \frac{MK}{\sum_{m=1}\sum_{k=1}\left(\left|r_{m,k}-\hat{z}_{m,k}\right|^2+\hat{\nu}^z_{m,k}\right)}.
		\label{b_beta2}
	\end{equation}
	
	Similar to (\ref{f_r2beta}), utilizing VMP rule, the forward message passing  from $f_{r_{m,k}}$ to $z_{m,k}$ can be given by
	\begin{equation}
	\begin{aligned}
			&m_{f_{r_{m,k}}\rightarrow z_{m,k}}(z_{m,k}) \propto\exp\left\{\int b(\beta)\ln f_{r_{m,k}}\mathrm{d}\beta\right\}\\
			\propto &\exp\left\{-\left|r_{m,k}-z_{m,k}\right|^2\int \beta b(\beta)\mathrm{d}\beta\right\}\\
			\overset{(a)}{\propto} &\mathcal{CN}(z_{m,k}; r_{m,k}, \hat{\beta}^{-1}), 
	\end{aligned}
	\label{m_fr2z}
	\end{equation}
	where $(a)$ is obtained by utilizing (\ref{b_beta2}).
	Owing to the Gaussian form of the message in (\ref{m_fr2z}), we can obtain the following model $\mathbf{r}_k = \mathbf{z}_k + \boldsymbol{\xi}_k$, where $\boldsymbol{\xi}_k \sim \mathcal{CN}(\boldsymbol{\xi}_k;0, \hat{\beta}^{-1}\mathbf{I}_{M})$, $\mathbf{r}_k = [r_{1,k}, r_{2,k}, \cdots, r_{M,k}]^{\mathrm{T}}$, and $\mathbf{z}_k = [z_{1,k}, z_{2,k}, \cdots, z_{M,k}]^{\mathrm{T}}$.
	This representation enables seamless integration with the measurement module of unitary AMP (UAMP) algorithm in \cite{GAMP5}.
	 
	Denote the posterior mean and variance of $\mathbf{x}_k$ as $\hat{\mathbf{x}}_k$ and $\hat{\nu}_{\mathbf{x}_k}$, which is defined in (\ref{post_mean}) and (\ref{post_variance}). According to the UAMP algorithm \cite{GAMP5}, define $\boldsymbol{\lambda}$ as $\left|\mathbf{A}\right|^2\mathbf{1}_{N}$. Thus, the message from $\mathbf{z}_k$ to $f_{\mathbf{r}_k}$ is given by $\mathcal{CN}(\mathbf{z}_k;\mathbf{p}_k, \boldsymbol{\nu}_{\mathbf{p}_k})$ with
	\begin{equation}
		\boldsymbol{\nu}_{\mathbf{p}_k} = \boldsymbol{\lambda}\hat{\nu}_{\mathbf{x}_k}, \quad \mathbf{p}_k = \mathbf{A}\hat{\mathbf{x}}_k - \boldsymbol{\nu}_{\mathbf{p}_k}\odot \boldsymbol{\mu}_k,
		\label{nu_pk} 
	\end{equation} 
	where $\odot$ denotes the element product; $\boldsymbol{\mu}_k$ is a intermediate vector, which is updated by
	\begin{align}
		\boldsymbol{\nu}_{\boldsymbol{\mu}_k} =\mathbf{1}_M./(\mathbf{1}_M\hat{\beta}^{-1}+\boldsymbol{\nu}_{\mathbf{p}_k}),
		\boldsymbol{\mu}_k = \boldsymbol{\nu}_{\boldsymbol{\mu}_k}\odot(\mathbf{r}_k-\mathbf{p}_k).
		\label{nu_muk}
	\end{align}
	
	In this case, utilizing SP rule, the belief of $\mathbf{z}_k$ is given by
	\begin{equation}
		\begin{aligned}
			b(\mathbf{z}_k) &= \mathcal{CN}(\mathbf{z}_k; \mathbf{p}_k, \boldsymbol{\nu}_{\mathbf{p}_k})\prod_{m=1} m_{f_{r_{m,k}}\rightarrow z_{m,k}}(z_{m,k})\\
			&=\mathcal{CN}(\mathbf{z}_k; \mathbf{p}_k, \boldsymbol{\nu}_{\mathbf{p}_k}) \mathcal{CN}(\mathbf{z}_k; \mathbf{r}_k, \hat{\beta}^{-1}\mathbf{I}_{M})\\
			&{\propto}\mathcal{CN}(\mathbf{z}_k; \hat{\mathbf{z}}_k, \boldsymbol{\nu}_{\mathbf{z}_k}),
		\end{aligned}
		\label{bz_m}
	\end{equation}
	where $\boldsymbol{\mu}_k$ and $\boldsymbol{\nu}_{\boldsymbol{\mu}_k}$ are respectively given by
	\begin{align}
		\label{nu_zk}{\boldsymbol{\nu}}_{\mathbf{z}_k} &=\boldsymbol{\nu}_{\mathbf{p}_k}./(\mathbf{1}+\hat{\beta}\boldsymbol{\nu}_{\mathbf{p}_k}), \\
		\label{nu_zk2}\hat{\mathbf{z}}_k &= (\mathbf{p}_k+\hat{\beta}\boldsymbol{\nu}_{\mathbf{p}_k}\odot\mathbf{r}_k) ./(\mathbf{1}+\hat{\beta}\boldsymbol{\nu}_{\mathbf{p}_k}).
	\end{align}
	
	Moreover, the output message from measurement module is given by 
	\begin{equation}
		\boldsymbol{\nu}_{\mathbf{q}_k} = \mathbf{1}./(\left|\mathbf{A}\right|^2\boldsymbol{\nu}_{\boldsymbol{\mu}_k}), \quad \mathbf{q}_k = \hat{\mathbf{x}}_k + \boldsymbol{\nu}_{\mathbf{q}_k} \odot (\mathbf{A}^{\mathrm{H}}\boldsymbol{\mu}_k).
		\label{nu_qk}
	\end{equation}
	\vspace{-3em}
	\subsection{Sparse Prior Module}
	The aim of the prior module is to update the distribution of ${x}_{n,k}$ according to extrinsic message $m_{x_{n,k} \rightarrow f_{x_{n,k}}}(x_{n,k})\propto \mathcal{CN}({x}_{n,k};{q}_{n,k}, {\nu}^{q}_{n,k})$ and calculate the posterior estimation $\hat{{x}}_{n,k}$ and $\hat{\nu}_{\mathbf{x}_k}$, where ${q}_{n,k}$ and ${\nu}^{q}_{n,k}$ denotes the $n$-th element of $\mathbf{q}_k$  and $\boldsymbol{\nu}_{\mathbf{q}_k}$, respectively.
	
	\subsubsection{Forward Message Passing} Denote the belief of $x_{n,k}$ as $b(x_{n,k}; \hat{x}_{n,k}, \hat{\nu}_{\mathbf{x}_k})$, defined in (\ref{b_xnk}). Utilizing VMP rule, similar to (\ref{f_r2beta}), the message from $f_{x_{n,k}}$ to $\gamma_{n,k}$ is given by
	\begin{equation}
		\begin{aligned}
			m_{f_{x_{n,k}} \rightarrow \gamma_{n,k}}(\gamma_{n,k}) \propto& \exp\left\{\int {b}(x_{n,k})\ln f_{x_{n,k}}\mathrm{d}x_{n,k}\right\}\\
			\propto& \gamma_{n,k}\exp(-\gamma_{n,k}(\left|\hat{x}_{n,k}\right|^2+\hat{\nu}_{\mathbf{x}_k})).
		\end{aligned}
		\label{m_fx2ga}
	\end{equation}   
	
	Furthermore, according to SP rule, the message from $f_{\gamma_{n,k}}$ to $s_{n,k}$ is given by
	\begin{equation}
		\begin{aligned}
			&m_{f_{\gamma_{n,k}} \rightarrow s_{n,k}}(s_{n,k}) 
			\overset{(a)}{\propto} \int f_{\gamma_{n,k}} m_{f_{x_{n,k}} \rightarrow \gamma_{n,k}}b(\boldsymbol{\alpha}_k)  \mathrm{d}\gamma_{n,k}\mathrm{d}\boldsymbol{\alpha}_k\\
			&=\pi^{\mathrm{out}}_{n,k}\delta(1-s_{n,k}) + (1-\pi^{\mathrm{out}}_{n,k})\delta(1+s_{n,k}),
		\end{aligned}
		\label{piout}
	\end{equation}
	where $(a)$ is obtained by utilizing the approximation $b(\boldsymbol{\alpha}_k)=b(\alpha^1_k)b(\alpha^2_k) \approx \prod_{j\neq n} m_{\boldsymbol{\alpha}_k\rightarrow f_{\gamma_{j,k}}}(\gamma_{j,k})$ and the expectation of Gamma distribution; $\pi^{\mathrm{out}}_n$ is defined as
	\begin{equation*}
		\begin{aligned}
			\pi^{\mathrm{out}}_{n,k} = 	\frac{\hat{a}_k^1(\hat{b}_k^2+\left|\hat{x}_{n,k}\right|^2+\hat{\nu}_{\mathbf{x}_k}))}{\hat{a}_k^1(\hat{b}_k^2+\left|\hat{x}_{n,k}\right|^2+\hat{\nu}_{\mathbf{x}_k}))+\hat{a}_k^2(\hat{b}_k^{1}+\left|\hat{x}_{n,k}\right|^2+\hat{\nu}_{\mathbf{x}_k})}.
		\end{aligned}
		\label{pi_out}
	\end{equation*}
	
	Utilizing the message $m_{f_{\gamma_{n,k}} \rightarrow s_{n,k}}(s_{n,k})$, we further derive the message update in the MRF with a 4-connect scheme, where $s_{n,k}^{{\text{l}}} \triangleq s_{n,k-1}$, 
	$s_{n,k}^{\text{r}} \triangleq s_{n,k+1}$, $s_{n,k}^{{\text{t}}} \triangleq s_{n-1,k}$, and $s_{n,k}^{{\text{b}}} \triangleq s_{n,k+1}$ denote the left, right, top and bottom neighbors of $x_{n,k}$. 
	The input message of $s_{n,k}$ from left, right, top and bottom neighbors, denoted as $m_{n,k}^{\mathrm{l}}$, $m_{n,k}^{\mathrm{r}}$, $m_{n,k}^{\mathrm{t}}$, and $m_{n,k}^{\mathrm{b}}$, are Bernoulli distributions.
	Take $m_{n,k}^{\mathrm{l}}$ as an example, according to SP rule, we have $m_{n,k}^{\mathrm{l}}
	\propto\lambda_{n,k}^{\mathrm{l}}\delta(1-s_{n,k}) + (1-\lambda_{n,k}^{\mathrm{l}})\delta(1+s_{n,k})$,
	where $\lambda_{n,k}^{\mathrm{l}}$ is given by (\ref{left}), as shown in the top of this page.
	\begin{figure*}[!t]
		\normalsize
		\setcounter{MYtempeqncnt}{\value{equation}}
		\begin{equation}
			\lambda_{n,j}^{\mathrm{l}} = \frac{\pi^{\mathrm{out}}_{s_{n,j}^{{\text{l}}}}\mathrm{e}^{\varpi-\eta}\prod_{w \in \left\{\mathrm{l,t,b}\right\}}\lambda^{w}_{s_{n,j}^{{\text{l}}}} + (1-\pi^{\mathrm{out}}_{s_{n,j}^{{\text{l}}}})\mathrm{e}^{\eta-\varpi}\prod_{w \in \left\{\mathrm{l,t,b}\right\}}(1-\lambda^{w}_{s_{n,j}^{{\text{l}}}})}{(\mathrm{e}^{\varpi}+ \mathrm{e}^{-\varpi})\left( \mathrm{e}^{-\eta}\pi^{\mathrm{out}}_{s_{n,j}^{{\text{l}}}}\prod_{w \in \left\{\mathrm{l,t,b}\right\}}\lambda^{w}_{s_{n,j}^{{\text{l}}}} + \mathrm{e}^{\eta}(1-\pi^{\mathrm{out}}_{s_{n,j}^{{\text{l}}}})\prod_{w \in \left\{\mathrm{l,t,b}\right\}}(1-\lambda^{w}_{s_{n,j}^{{\text{l}}}})\right)}.
			\label{left}
		\end{equation}
		\hrulefill
		\vspace*{4pt}
	\end{figure*}
	The other messages can be obtained in a similar way.
	
	\subsubsection{Backward Message Passing}
	With the messages of neighbors and $v(s_{n,k})$ and SP rule, the message from $s_{n,k}$ to $ f_{\gamma_{n,k}}$ can be given by 
	\begin{equation}
		\begin{aligned}
			&m_{s_{n,k}\rightarrow f_{\gamma_{n,k}}} = \prod_{w\in\left\{\mathrm{l,r,t,b}\right\}} m_{n,k}^{\mathrm{l}}v(s_{n,k})\\
			\propto& \pi^{\mathrm{in}}_{n,k} \delta(1-s_{n,k}) + (1-\pi^{\mathrm{in}}_{n,k}) \delta(1+s_{n,k}),
		\end{aligned}
	\end{equation}
	where $\pi^{\mathrm{in}}_{n,k}$ is defined as
	\begin{equation}
		\frac{\mathrm{e}^{-\eta}\prod_{w\in\left\{\mathrm{l,r,t,b}\right\}}\lambda_{n,k}^{w}}{\mathrm{e}^{-\eta}\prod_{w\in\left\{\mathrm{l,r,t,b}\right\}}\lambda_{n,k}^{w}+\mathrm{e}^{\eta}\prod_{w\in\left\{\mathrm{l,r,t,b}\right\}}(1-\lambda_{n,k}^{w})}.
		\label{pi_in}
	\end{equation}
	
	Using SP rule, the message from $f_{\gamma_{n,k}}$ to $\alpha_k^1$ is given by
	\begin{align}
		\notag m_{f_{\gamma_{n,k}} \rightarrow \alpha_k^1}(\alpha_k^1) \propto& \int f_{\gamma_{n,k}} m_{f_{x_{n,k}} \rightarrow \gamma_{n,k}}  m_{s_{n,k}\rightarrow f_{\gamma_{n,k}}} \mathrm{d}\gamma_{n,k} \mathrm{d}s_{n,k},\\
		\overset{(a)}{\propto}&\alpha_k^1\exp(-\alpha_k^1(\left|\hat{x}_{n,k}\right|^2+\hat{\nu}_{\mathbf{x}_k})),
	\end{align}
	where $(a)$ is obtained by utilizing the property of delta function.
	Similarly, $m_{f_{\gamma_{n,k}} \rightarrow \alpha_k^2}(\alpha_k^2)
	\propto \alpha_{k}^2\exp(-\alpha_{k}^2(\left|\hat{x}_{n,k}\right|^2+\hat{\nu}_{\mathbf{x}_k}))$.
	As a result, the belief of $\alpha_k^1$ is given by
	\begin{equation}
		{b}(\alpha^1_k) \propto p(\alpha_k^1) \prod_{n}m_{f_{\gamma_{n,k}} \rightarrow \alpha_k^1} (\alpha_k^1)
		\propto\mathrm{Ga}(\alpha_k^1; \hat{a}_k^1, \hat{b}_k^1),
		\label{bk1}
	\end{equation}
	where $\hat{a}_k^1 = a_1+N$ and $\hat{b}_k^1 = b_1+\sum_{n=1}^N\left|\hat{x}_{n,k}\right|^2+\hat{\nu}_{\mathbf{x}_k}$. Similarly, we have ${b}(\alpha^2_k) 
	\propto\mathrm{Ga}(\alpha_k^2; \hat{a}_k^2, \hat{b}_k^2)$ with $\hat{a}_k^2 = a_2+N$ and $\hat{b}_k^2 = b_2+\sum_{n=1}^N\left|\hat{x}_{n,k}\right|^2+\hat{\nu}_{\mathbf{x}_k}$. According to SP rule, the message from $f_{\gamma_{n,k}}$ to $\gamma_{n,k}$ is given by
	\begin{equation}
		\begin{aligned}
			m_{f_{\gamma_{n,k}} \rightarrow \gamma_{n,k}}(\gamma_{n,k})= \pi^{\mathrm{in}}_{n,j}{b}(\alpha^1_k) + (1-\pi^{\mathrm{in}}_{n,j}){b}(\alpha^2_k).
		\end{aligned}
		\label{f_ga2ga}
	\end{equation}
	
	Consequently, the belief of $\gamma_{n,k}$ is given by
	\begin{equation}
		\begin{aligned}
			&b(\gamma_{n,k}) \propto m_{f_{\gamma_{n,k}} \rightarrow \gamma_{n,k}}(\gamma_{n,k})m_{f_{x_{n,k}} \rightarrow \gamma_{n,k}}(\gamma_{n,k})\\
			\propto& \pi^{\mathrm{in}}_{n,j}\gamma_{n,k}^{\hat{a}_{k}^1}\exp(-\gamma_{n,k}(\hat{b}_{k}^1+\left|\hat{x}_{n,k}\right|^2+\hat{\nu}_{\mathbf{x}_k}))\\
			+&(1-\pi^{\mathrm{in}}_{n,k})\gamma_{n,k}^{\hat{a}_{k}^2}\exp(-\gamma_{n,k}(\hat{b}_{k}^2+\left|\hat{x}_{n,k}\right|^2+\hat{\nu}_{\mathbf{x}_k})).
		\end{aligned}
	\end{equation}
	
	As a result, utilizing VMP rule, the message from $f_{x_{n,k}}$ to $x_{n,k}$ is denoted as 
	\begin{equation}
	\begin{aligned}
			&m_{f_{x_{n,k}} \rightarrow x_{n,k}}(x_{n,k})\propto \exp\left\{\int b(\gamma_{n,k})\ln f_{x_{n,k}}\mathrm{d}\gamma_{n,k}\right\} \\
			&\propto \exp\left\{-\int\left|x_{n,k}\right|^2\gamma_{n,k} b(\gamma_{n,k})\mathrm{d}\gamma_{n,k}\right\} \\
			&\overset{(a)}{\propto}\mathcal{CN}(x_{n,k}; 0, \hat{\gamma}_{n,k}^{-1}),
	\end{aligned}
	\end{equation}
	 where $(a)$ is obtained by 
	\begin{equation}
		\begin{aligned}
			\hat{\gamma}_{n,k} &= \int \gamma_{n,k}b(\gamma_{n,k})\mathrm{d}\gamma_{n,k}
			= \pi^{\mathrm{in}}_{n,k}\frac{\hat{a}_{k}^1+1}{\hat{b}_{k}^1+\left|\hat{x}_{n,k}\right|^2+\hat{\nu}_{\mathbf{x}_k}}\\&+ (1-\pi^{\mathrm{in}}_{n,k})\frac{\hat{a}_{k}^2+1}{\hat{b}_{k}^2+\left|\hat{x}_{n,k}\right|^2+\hat{\nu}_{\mathbf{x}_k}}.
		\end{aligned}
		\label{hat_ga}
	\end{equation}
	
	Combining the message from measurement module as $\mathcal{CN}({x}_{n,k};{q}_{n,k}, \nu_{n,k}^q)$. Thus, the approximate posterior distribution of $x_{n,j}$ can be approximated as
	\begin{equation}
		\begin{aligned}
			b(x_{n,k}) &\propto \mathcal{CN}({x}_{n,k};{q}_{n,k}, \nu_{n,k}^q)\mathcal{CN}(x_{n,k}; 0, \hat{\gamma}_{n,k}^{-1})\\
			&\overset{(a)}{\propto} \mathcal{CN}(x_{n,k}; \hat{x}_{n,k}, \hat{\nu}_{\mathbf{x}_k}),
		\end{aligned}
		\label{b_xnk}
	\end{equation}
	where $(a)$ is obtained similar to (\ref{bz_m}), and the approximate posterior mean and variance of $x_{n,k}$ are respectively given by
	\begin{equation}
		\hat{\nu}_{n,k}^{x} = \frac{\nu_{n,k}^q}{1+\nu_{n,k}^q\hat{\gamma}_{n,k}}, \quad
		\hat{x}_{n,k} = \frac{q_{n,k}}{1+\nu_{n,k}^q\hat{\gamma}_{n,k}}.
		\label{post_mean}
	\end{equation}
	Performing the average operations to $\nu_{n,k}^{x}$, we further have 
	\begin{equation}
		\hat{\nu}_{\mathbf{x}_k} = \frac{1}{N}\sum_{n=1}^{N}\hat{\nu}_{n,k}^{x}.
		\label{post_variance}
	\end{equation}
	
	\begin{algorithm}
		\renewcommand{\algorithmicrequire}{\textbf{Input:}}
		\renewcommand{\algorithmicensure}{\textbf{Output:}}
		\caption{Proposed MMV-HMP algorithm}
		\begin{algorithmic}[1]
			\Require received vector $\mathbf{R}$, measurement matrix $\mathbf{A}$.
			\Statex \textbf{Initialize:} $\hat{\nu}_{\mathbf{x}_k}^{(0)}=1$, $\hat{\mathbf{x}}_k=\mathbf{0}$, $\hat{{\gamma}}_{n,k}={1}, \hat{\beta} = 1$, and $\boldsymbol{\mu}_k=\mathbf{0}$. 
			\While{the stopping criterion is not met}
			\Statex /*\textbf{Measurement module}*/
			\State Update the $\boldsymbol{\nu}_{\mathbf{p}_k}$ and $\mathbf{p}_k$ according to (\ref{nu_pk});
			\State Update the $\boldsymbol{\nu}_{\boldsymbol{\mu}_k}$ and $\boldsymbol{\mu}_k$ according to (\ref{nu_muk});
			\State Update the $\boldsymbol{\nu}_{\mathbf{z}_k}$ and $\mathbf{z}_k$ according to (\ref{nu_zk}) and (\ref{nu_zk2});
			\State Update $\hat{\beta}$ according to (\ref{b_beta2});
			\State Update the $\boldsymbol{\nu}_{\mathbf{q}_k}$ and $\mathbf{q}_k$ according to (\ref{nu_qk});
			\Statex /*\textbf{Prior Module}*/
			\State Update the messages $m_{f_{x_{n,k}} \rightarrow s_{n,k}}$ according to (\ref{piout});
			\State Update the messages $m_{n,k}^{\mathrm{l}}$, $m_{n,k}^{\mathrm{r}}$, $m_{n,k}^{\mathrm{t}}$, and $m_{n,k}^{\mathrm{b}}$;
			\State Update the messages $m_{s_{n,k}\rightarrow f_{\gamma_{n,k}}}$ according to (\ref{pi_in});
			\State Update the belief $b(\alpha_k^1)$ and $b(\alpha_k^2)$ according to (\ref{bk1});
			\State Update $\hat{\gamma}_{n,k}$ according to (\ref{hat_ga});
			\State Update $\hat{x}_{n,k}$ and $\hat{\nu}_{\mathbf{x}_k}$ according to (\ref{post_mean}) and (\ref{post_variance}).
			\EndWhile
			\Ensure $\hat{x}_{n,k}$.
		\end{algorithmic}
		\label{UAMP_SBL_MRF}
	\end{algorithm}
	
	The proposed MMV-HMP algorithm is summarized in Algorithm \ref{UAMP_SBL_MRF} and it can be terminated when it reached a maximum number of iteration or the difference between the estimates of two consecutive iterations is less than $10^{-5}$.
	In the following, we provide the computational complexity analysis for the proposed MMV-HMP algorithm. Examining the steps of Algorithm~\ref{UAMP_SBL_MRF}, it is evident that there is no matrix inversion involved. Thus, the most computationally intensive parts only involve matrix-vector products in lines 2 and 6, i.e., $\mathcal{O}(MN)$ per iteration. Consequently, the total complexity of the MMV-HMP algorithm is $\mathcal{O}(TKMN)$, where $T$ denotes the number of iterations.
	\vspace{-1em}
	\section{Simulation Results}
	\label{section6}
	\begin{table}
		\renewcommand\arraystretch{1.1}
		\centering
		\caption{Simulation Parameters}
		\setlength{\tabcolsep}{4mm}{
			\begin{tabular}{c c}
				\toprule [1pt]
				\makecell[l]{Notations}& Parameters\\ 
				\midrule [0.5pt]
				\makecell[l]{Number of BS antenna $N_{\mathrm{R}}$} &256 \\
				\makecell[l]{Number of RF chain $N_{\mathrm{RF}}$} &16 \\
				\makecell[l]{Carrier frequency $f_c$}    &30GHz\\
				\makecell[l]{Number of pilot carriers $K$} &64 \\
				\makecell[l]{System bandwidth $f_s$} &1.6GHz \\
				\makecell[l]{Number of channel path $L$} &4\\
				\makecell[l]{Angle of arrival $\vartheta_l$} &$\mathcal{U}(-\pi/2, \pi/2)$\\
				\makecell[l]{Distance between BS and UE or scatterers $r_l$} &[5, 50]m\\
				\makecell[l]{Proportion of visible antenna elements $\rho_l$} &(0,1]\\
				\bottomrule[1pt]
			\end{tabular}
		}
		\label{Parameters}
	\end{table}
	In this section, we evaluate the performance of the proposed channel estimation scheme under various system setups. The simulation parameters are shown in Table \ref{Parameters}. 
	In particular, we consider normalized mean square error (NMSE) as performance metrics, which is defined as $\mathrm{NMSE} \triangleq {\lVert \hat{\mathbf{H}} - \mathbf{H} \rVert^2_{\mathrm{F}}}/{\lVert \mathbf{H} \lVert^2_{\mathrm{F}}}$,
	where $\mathbf{H}$ and $\hat{\mathbf{H}}$ are the true channel and estimated channel, respectively. In addition, the SNR is defined in received side, which is given by $10\log_{10}\left(\lVert\mathbf{WH} \rVert^2_{\mathrm{F}}/ \lVert\mathbf{N} \rVert^2_{\mathrm{F}}\right)$.
	Additionally, we compare the proposed MMV-HMP algorithm
	with the following baselines:
	\begin{itemize}
		\item \textbf{SOMP}\cite{PolarCS}: 
		The simultaneous OMP algorithm designed for on-grid sparse recovery, which depends on the knowledge of the number of non-zeros components.
		\item \textbf{StdSBL}\cite{StdSBL}: The standard SBL algorithm, implemented within an expectation-maximization (EM) framework. The algorithm employs a two-layer Gaussian-Gamma hierarchical prior model, where the posterior estimates of $\mathbf{x}_k$ are updated in the E-step, and the hyperparameters of the prior model are updated in the M-step.
		\item \textbf{UAMP-SBL}\cite{GAMP5}: An improved version of the StdSBL algorithm that leverages the  UAMP framework to perform the E-step and variational message passing to achieve the updates of prior parameters.
		\item \textbf{PC-SBL}\cite{PC_SBL}: A variant of the StdSBL framework that incorporates a pattern-coupled Gaussian prior model to exploit the block sparsity inherent in signals.
		\item \textbf{VSP}\cite{Variance_State_1}: A variant of the StdSBL framework that employs a MRF-based hierarchical prior model to effectively capture the block sparsity of signals. Unlike the StdSBL and PC-SBL methods, the prior parameters in the VSP algorithm are updated using a moment-matching approach, providing a computationally efficient alternative.
	\end{itemize}
	
	\begin{figure}
		\centering
		\includegraphics[width=0.4\textwidth]{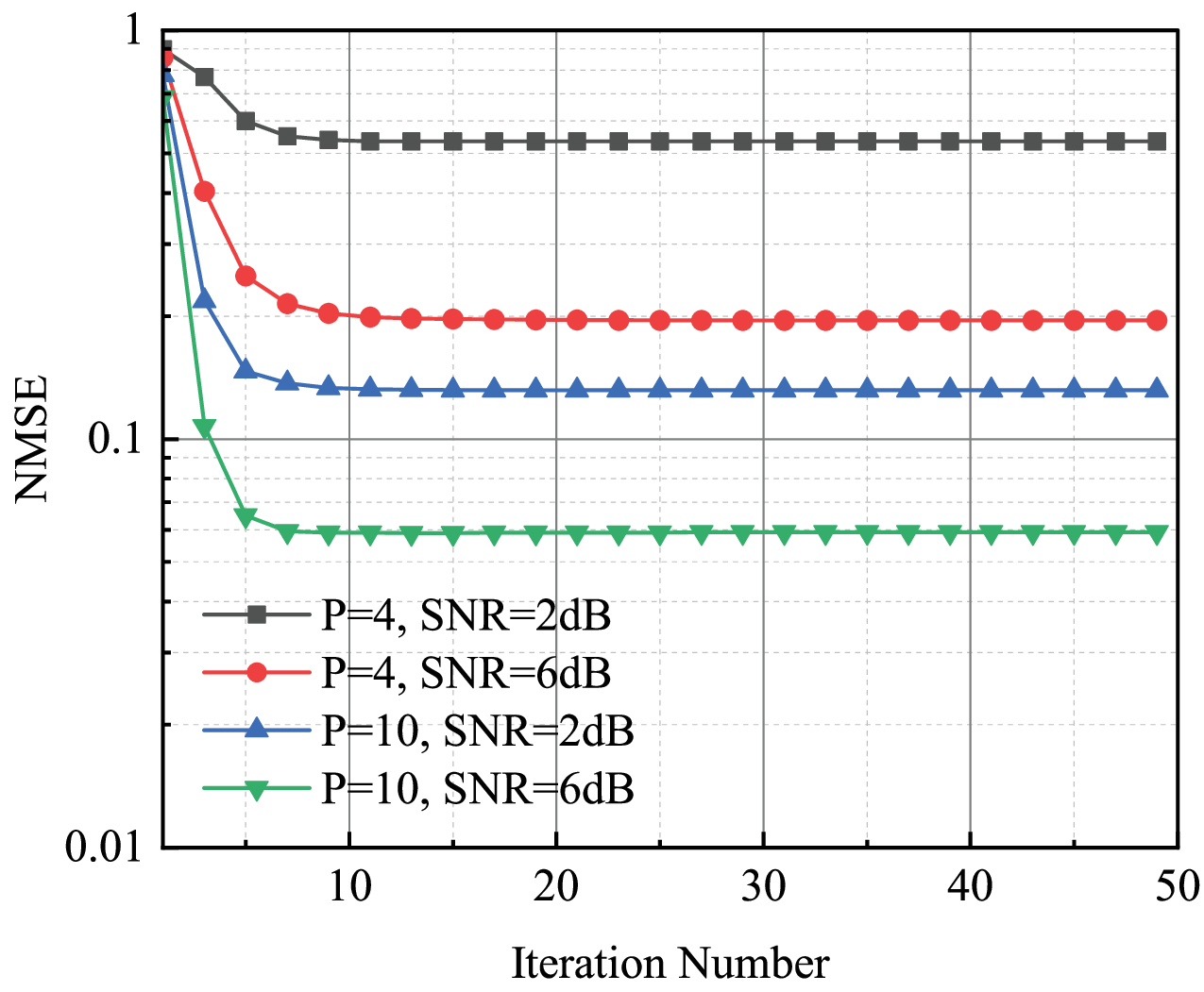}
		\caption{Convergence behavior of MMV-HMP.}
		\label{Convergence_Performance}
	\end{figure}
	\begin{figure}
		\centering
		\includegraphics[width=0.4\textwidth]{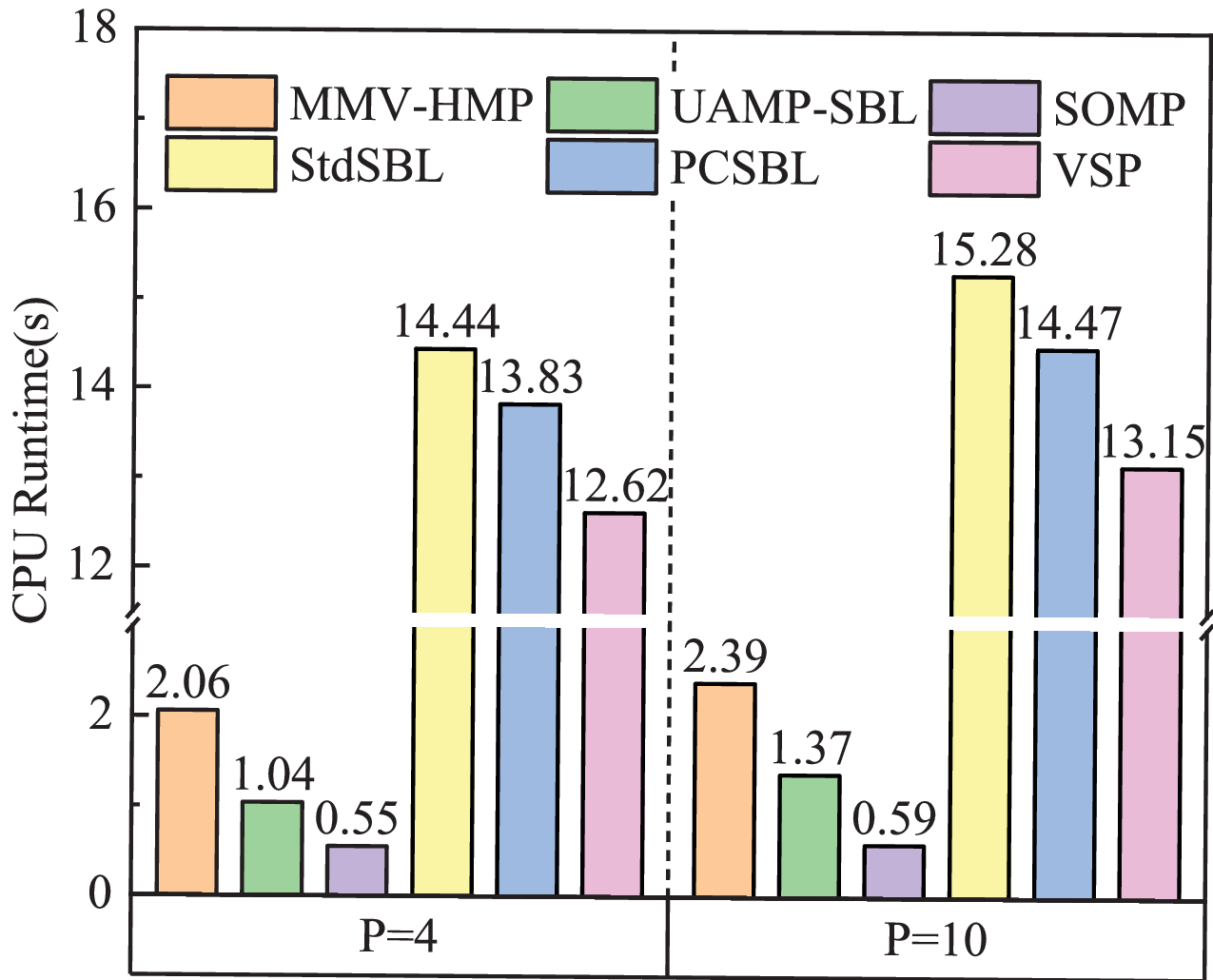}
		\caption{Complexity evaluation of different algorithms.}
		\label{Runtime}
	\end{figure}
	\vspace{-1em}
	\subsection{Convergence and Complexity Performance}
	Fig. \ref{Convergence_Performance} illustrates the convergence behavior of the proposed MMV-HMP algorithm by plotting the NMSE against the number of iterations under various pilot lengths and SNR conditions. The results clearly demonstrate a consistent monotonic decrease in NMSE across all iterations, indicating the algorithm's stable and reliable convergence across different simulation scenarios. Based on the trade-off between estimation accuracy and computational complexity, it is observed that the NMSE stabilizes after approximately 20 iterations. Therefore, for subsequent simulations, the maximum number of iterations can be effectively limited to 20 without compromising performance.
	
	Fig. \ref{Runtime} presents a computational complexity analysis by comparing the central processing unit (CPU) runtime of various algorithms. Among these, SOMP exhibits the shortest runtime. This efficiency is attributed to its computational simplicity, involving only matrix-vector product between the residual and the measurement matrix, as well as a projection operation between the received signal and the low-dimensional basis matrix. In contrast, SBL-based methods such as StdSBL, PC-SBL, and VSP require significantly more computational time. This is primarily due to the matrix inversion operations inherent in their implementation, which increase computational complexity. As anticipated, the proposed MMV-HMP algorithm demonstrates a shorter runtime compared to StdSBL, PC-SBL, and VSP due to avoidance of matrix inversion operations and stable convergence properties. 
	\vspace{-1em}
	\subsection{NMSE versus SNR and  Pilot Symbol Number}
\begin{figure}
	 	\centering
	 	\includegraphics[width=0.4\textwidth]{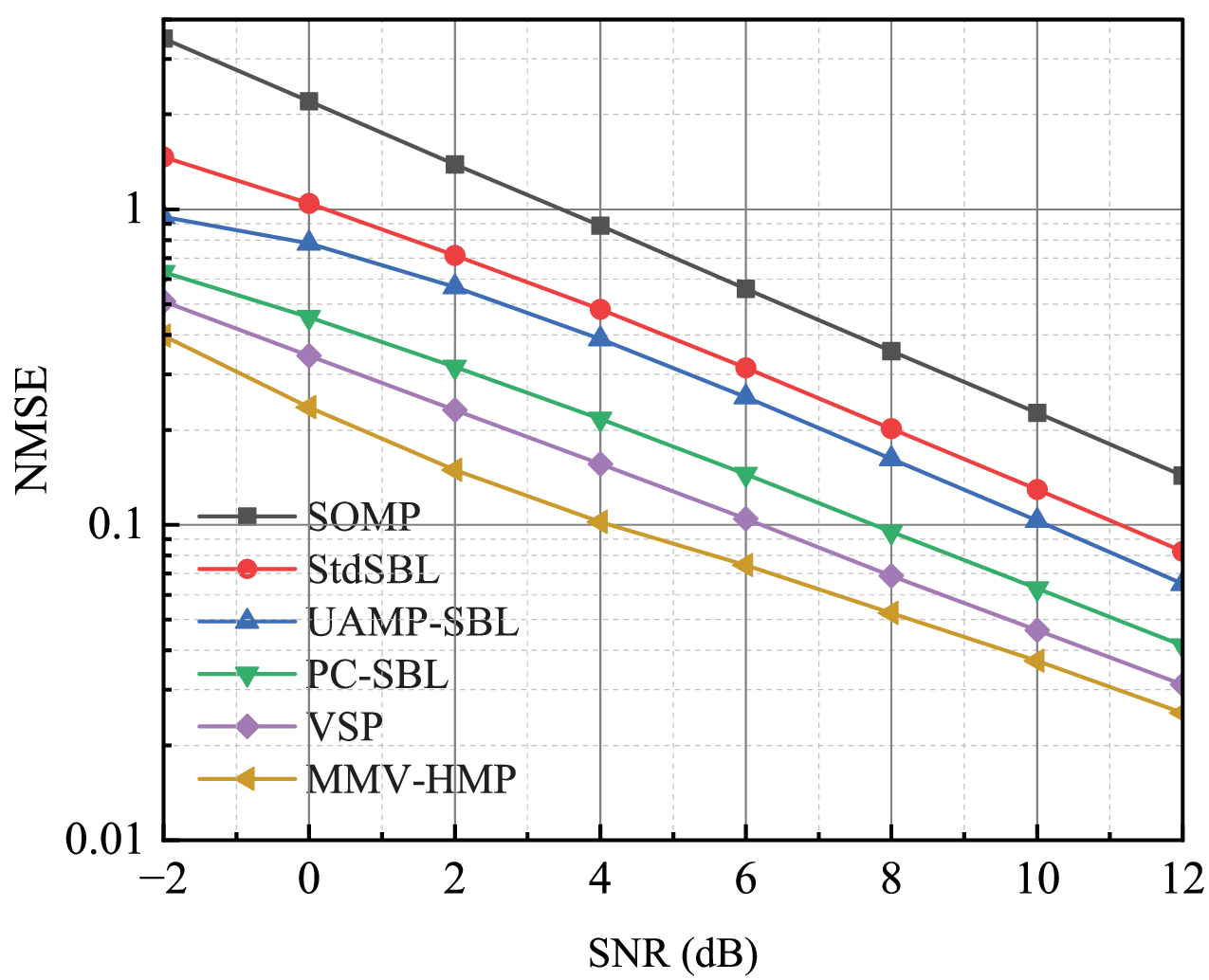}
	 	\caption{NMSE versus SNR.}
	 	\label{SNR_Performance}
	 \end{figure}

	 Fig. \ref{SNR_Performance} illustrates the NMSE performance of various algorithms as a function of SNR for $P=8$. The results reveal that algorithms such as SOMP \cite{OMP_1}, StdSBL \cite{StdSBL}, and UAMP-SBL \cite{GAMP5}, which fail to account for both global block sparsity and local common-delay sparsity, exhibit significantly poorer NMSE performance compared to algorithms like PC-SBL, VSP, and the proposed MMV-HMP.
	 Among block-sparsity prior-based methods, the PC-SBL algorithm leverages a pattern-coupled Gaussian prior model to effectively capture local common-delay sparsity. However, it does not consider global block sparsity. In contrast, the VSP algorithm employs a MRF-based prior to model global block sparsity but lacks the ability to incorporate local common-delay sparsity. Overall, the performance degradation of these SBL-based baselines stems from the mismatch between their prior models and the structured sparsity of XL-MIMO channels.
	 In contrast, the proposed MMV-HMP algorithm overcomes these limitations by utilizing a tailored column-wise hierarchical prior that simultaneously incorporates both MRF structure and precision sharing mechanism. This comprehensive modeling approach ensures consistently superior NMSE performance across the entire SNR range under consideration.
	 
	 \begin{figure}
	 	\centering
	 	\includegraphics[width=0.4\textwidth]{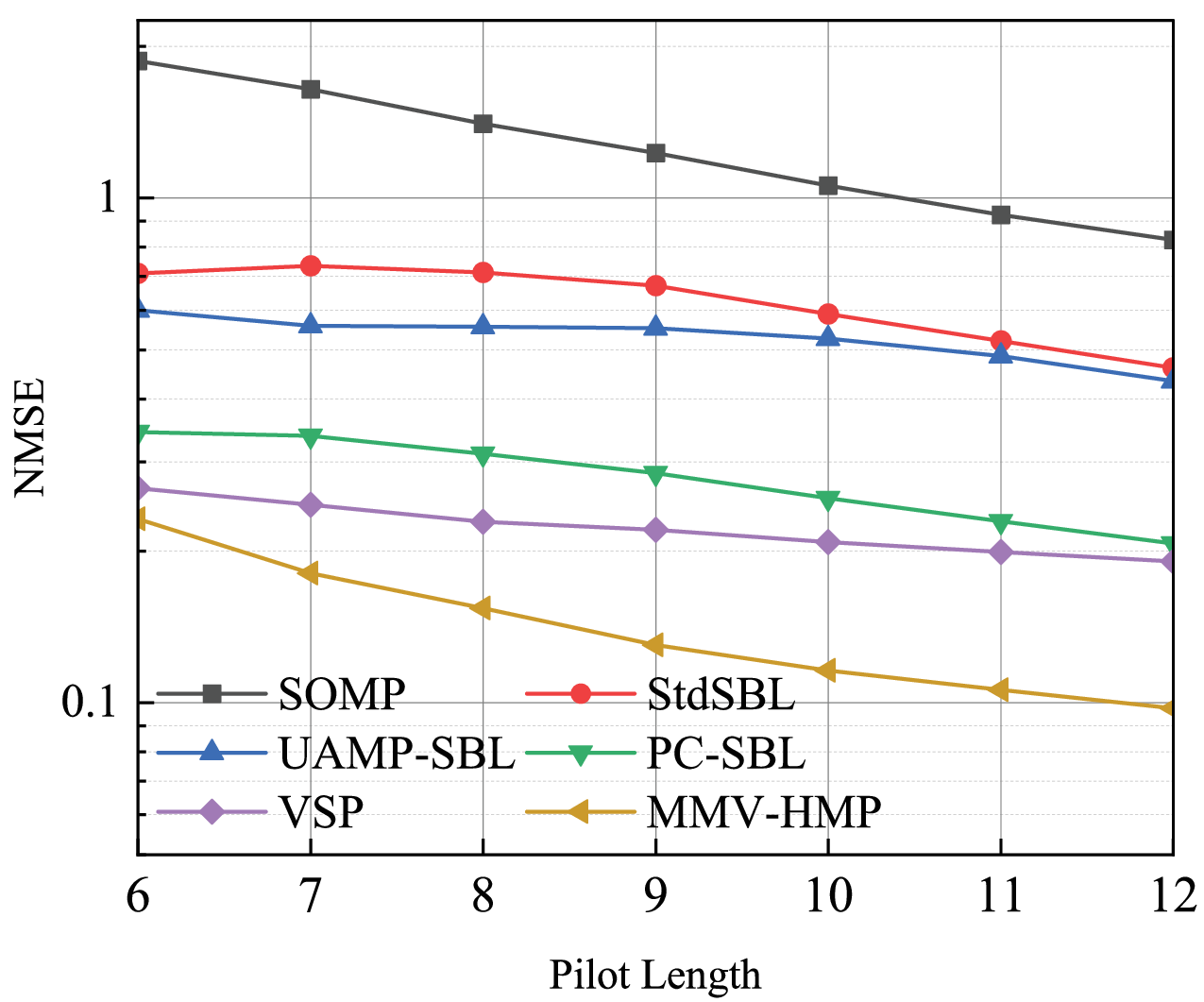}
	 	\caption{NMSE versus pilot symbol number.}
	 	\label{Pilot_Performance}
	 \end{figure} 
	 
	 Fig. \ref{Pilot_Performance} evaluates the estimation performance of various algorithms under different compression ratios by varying the number of pilot symbols, $P$, with the fixed SNR of 2 dB. As the number of pilot symbols varies from 6 to 12, the compression ratio $M/N$ ranges from 0.375 to 0.75. 
	 We can obtain two key observations: 1) Across most compression ratios, the proposed MMV-HMP algorithm outperforms baseline methods. This result highlights its robust sparse recovery capability under varying compression conditions. 
	 Moreover, the superior performance demonstrates that the proposed algorithm is a low-overhead solution, requiring fewer pilot symbols to achieve comparable or better performance than other algorithms.
	 2) As the pilot length increases, the rate of performance improvement gradually diminishes, eventually approaching saturation. This observation suggests that selecting a moderate value of $P$ is sufficient to strike a balance between estimation performance and pilot overhead.
	 \vspace{-1em}
	\subsection{NMSE versus Path Number}
	\begin{figure}
		\centering
		\includegraphics[width=0.4\textwidth]{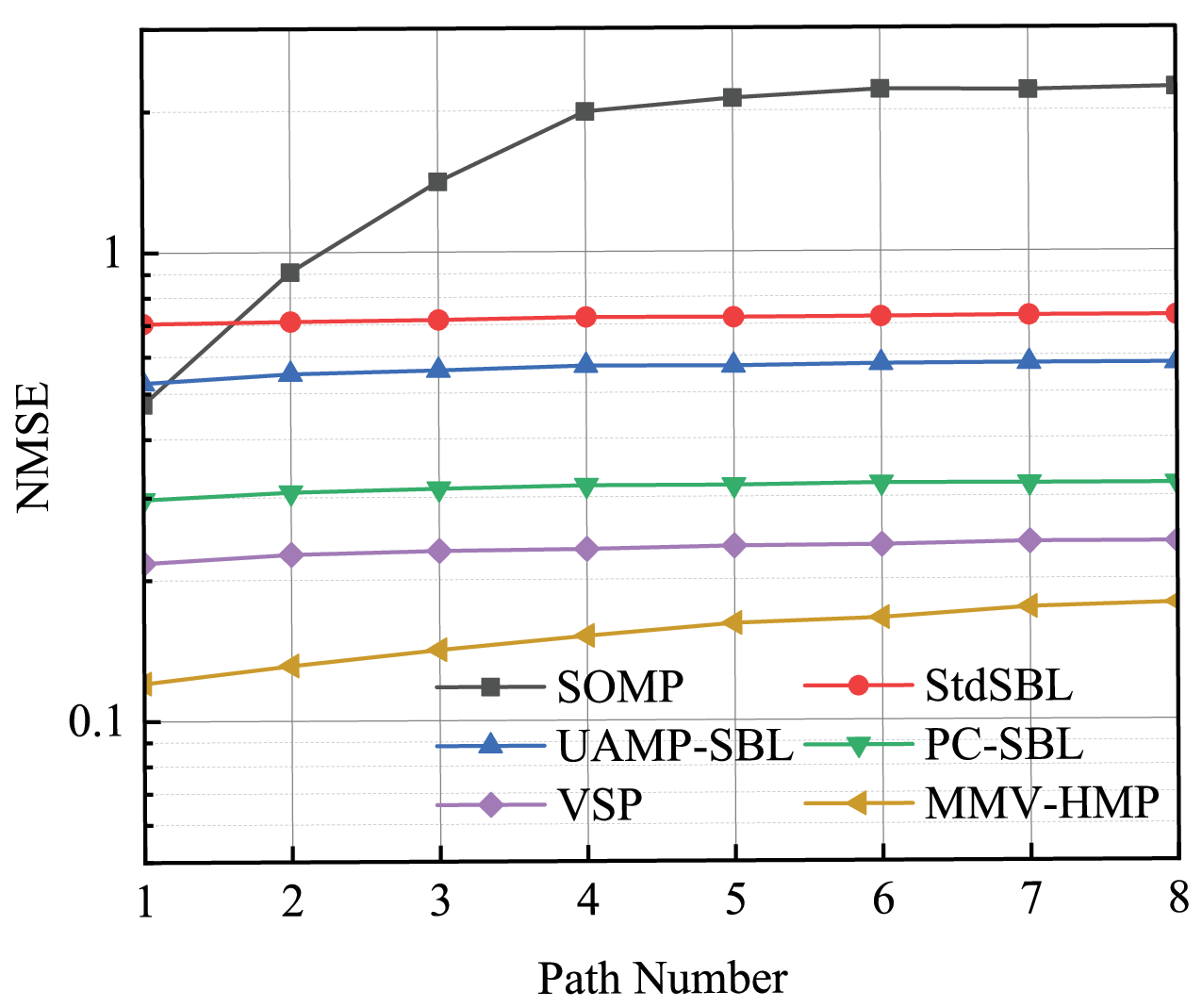}
		\caption{NMSE versus path number.}
		\label{Path_Performance}
	\end{figure} 
	Fig. \ref{Path_Performance} illustrates the NMSE performance of various algorithms as a function of the number of propagation paths, with $P=8$ and $\mathrm{SNR} =2$dB. The SOMP algorithm exhibits significant sensitivity to prior knowledge of the number of paths, with its performance degrading sharply when the assumed and actual path numbers differ.
	In contrast, Bayesian inference-based methods, leveraging sparsity-promoting prior models, adapt effectively to variations in the number of paths, maintaining consistent and robust performance across all scenarios. Notably, the proposed MMV-HMP algorithm consistently outperforms other Bayesian methods, demonstrating superior capability in handling dynamic scattering environments.
	
	\begin{figure}
		\centering
		\includegraphics[width=0.4\textwidth]{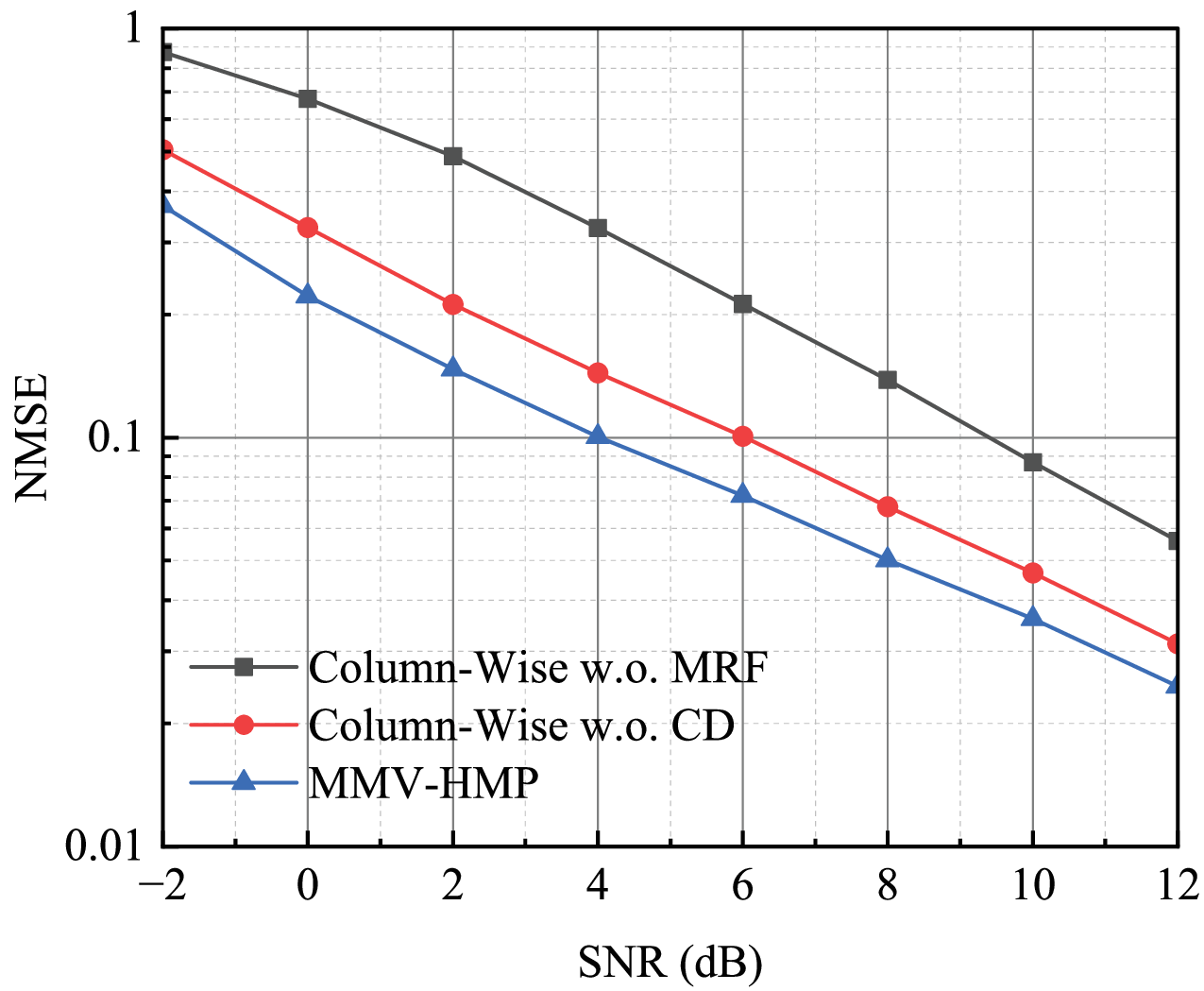}
		\caption{Ablation study of the column-wise hierarchical model.}
		\label{Prior_evaluation}
	\end{figure}
	
	\begin{figure*}[h]
		\centering
		\subfigure[Original.]{
			\includegraphics[width=0.4\textwidth]{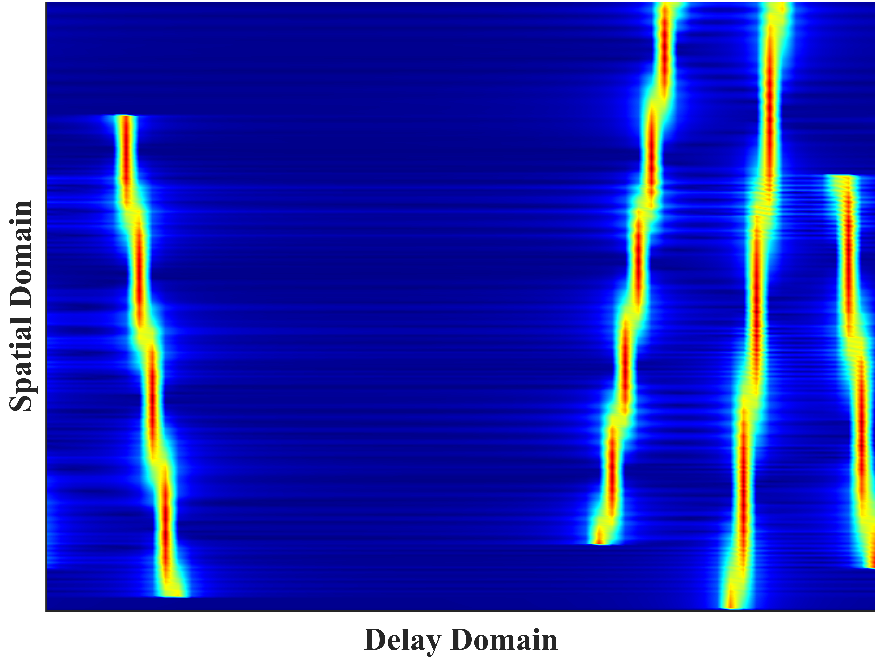}
			\label{original2}
		}
		\qquad
		\subfigure[Column-Wise w.o. MRF.]{
			\includegraphics[width=0.4\textwidth]{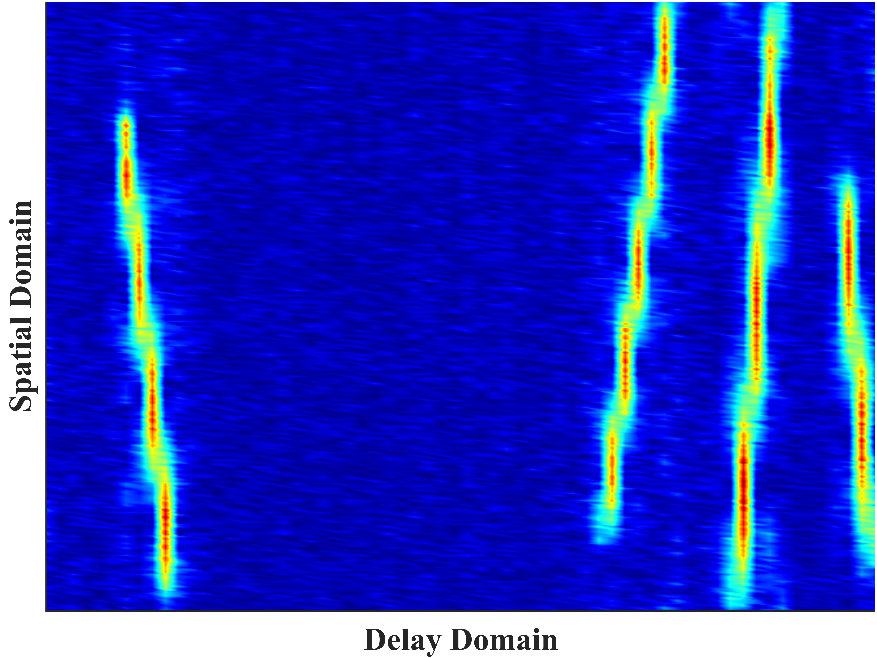}
			\label{es12}
		}
		
		\subfigure[Column-Wise w.o. CD.]{
			\includegraphics[width=0.4\textwidth]{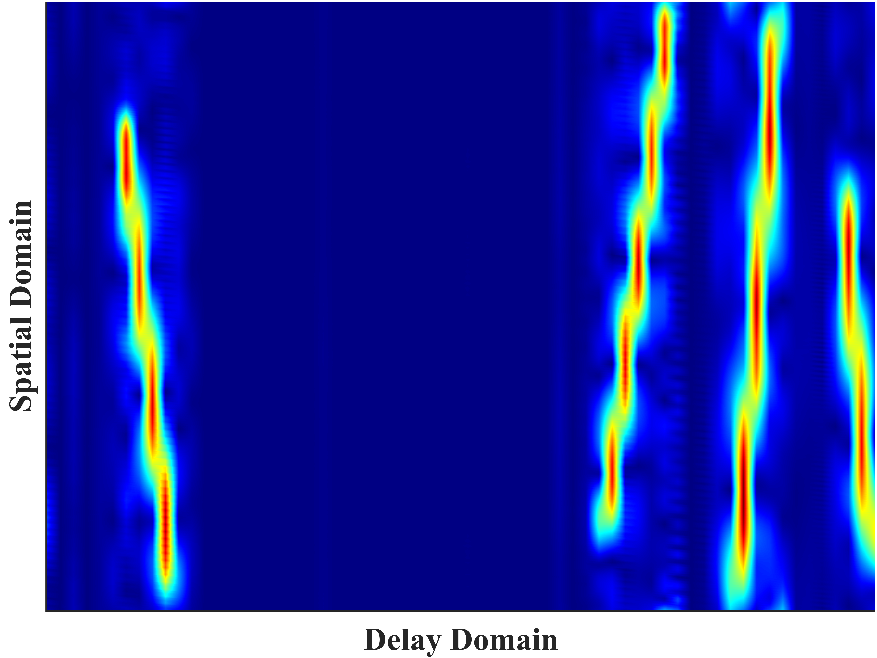}
			\label{es42}
		}
		\qquad
		\subfigure[MMV-HMP.]{
			\includegraphics[width=0.4\textwidth]{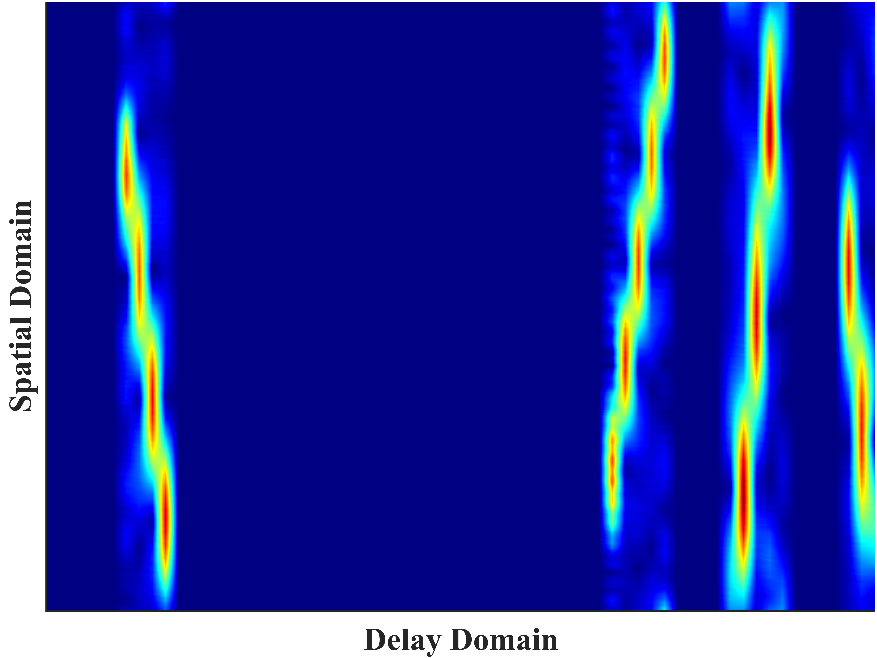}
			\label{es52}
		}
		\caption{Original and reconstructed channels by ``Column-Wise w.o. MRF,'' ``Column-Wise w.o. CD,'' and MMV-HMP algorithm.}
		\label{Simulation2}
	\end{figure*}

	\subsection{Ablation Study to Evaluate the Effects of Prior Models}
	To evaluate the effectiveness of the proposed column-wise hierarchical sparse prior model, we conduct an ablation study. Specifically, the proposed MMV-HMP algorithm is compared against the following two baselines:
	\begin{itemize}
		\item \textbf{Column-Wise w.o. MRF}: This baseline removes the MRF structure from the proposed column-wise hierarchical sparse prior model, thereby ignoring coefficient correlations.
		\item \textbf{Column-Wise w.o. CD}: This baseline eliminates the precision parameter sharing strategy in the proposed prior model, thus modeling each coefficient's precision independently without promoting local clustering.
	\end{itemize}
	
	Fig. \ref{Prior_evaluation} investigates the NMSE performance versus SNR for different prior models. As expected, the estimation performance of the two baselines exhibits significant degradation, as they fail to fully capture both the global and local structured sparsity. This validates the necessity of jointly employing the MRF structure and the precision parameter sharing mechanism.
	Moreover, it is observed that the performance degradation of the ``Column-Wise w.o. MRF'' model is substantially greater than that of the ``Column-Wise w.o. CD'' model. This indicates that the improvement introduced by incorporating the MRF is more critical than the gain achieved by the precision parameter sharing mechanism.
	
	To intuitively demonstrate the estimation performance of different priors, Fig.~\ref{Simulation2} presents the results of a Monte Carlo simulation with $\mathrm{SNR}=2$ dB and $P=8$. The original spatial-delay channels are compared with the reconstructed channels obtained using the baseline methods and the proposed MMV-HMP algorithm.
	The figure clearly shows that the proposed MMV-HMP algorithm achieves the best reconstruction quality, accurately recovering the spatial-delay channels and demonstrating superior sparse recovery capability. As expected, ``Column-Wise w.o. MRF'' yields the poorest reconstruction, with significant residual noise, mainly due to its limited ability to capture the global block sparsity.
	Furthermore, compared to ``Column-Wise w.o. CD,'' the channels reconstructed by MMV-HMP exhibit noticeably reduced power dispersion near the non-zero regions. 
	This improvement is attributed to the incorporation of the precision parameter sharing mechanism.
	\vspace{-1em}
	\section{Conclusions}
	\label{section7}
	In this paper, we have addressed the channel estimation problem in XL-MIMO systems, taking into account the spherical wavefront effects, SnS properties, and dual-wideband effects. We began by rigorously quantifying the angular and delay spread properties of SnS dual-wideband channels in the angular-delay domain and revealed their inherent global block sparsity and local common-delay sparsity. To exploit this structured sparsity, we proposed a computationally efficient MMV-HMP algorithm. Simulation results demonstrate the superiority of the MMV-HMP algorithm in both computational complexity and estimation performance. Furthermore, an ablation study confirms the effectiveness of the proposed column-wise hierarchical prior, validating its key contribution to enhanced estimation accuracy.
	
	\vspace{-1em}
	\section*{Appendix A Proof of Lemma 1}
	\setcounter{equation}{0}
	\renewcommand\theequation{A.\arabic{equation}} The proof of Lemma 1 is divided into two parts. First, we demonstrate that the angular-frequency representation $\mathbf{F}_{\mathrm{A}}^{\mathrm{H}}\boldsymbol{\Theta}(\varphi_l,\psi_l)$ exhibits row-wise sparsity. 
	Next, we show that the spatial-delay representation $\boldsymbol{\Theta}(\varphi_,\psi_l)\mathbf{F}_{\mathrm{D}}^*$ exhibits column-wise sparsity.
	
	For a fixed $k$, the angular transformation of $\boldsymbol{\theta}(f_k,\varphi_l,\psi_l)$ is given by
	\begin{equation}
		[\boldsymbol{\Xi}_{l,\mathrm{A}}]_{m,k} = \sum_{n=1}^N \mathrm{e}^{\mathrm{j}\frac{2\pi}{N}mn} \mathrm{e}^{\mathrm{j}2\pi \left(a_{l,k}n- \frac{b_{l,k}}{2}n\right)n}.
	\end{equation}
	
	According to (\ref{chirp2}) and (\ref{chirp3}), it can be obtained that $[\boldsymbol{\Xi}_{l,\mathrm{A}}]_{:,k}$ demonstrates block sparsity with its prominent spatial frequency components concentrated from $i_{l,e}(k)$ to $i_{l,s}(k)$.
	Extending this observation across all $K$ subcarriers, the overall spatial frequency range of significant components spans from $i_{l,e}^{\mathrm{min}} = \min_k i_{l,e}(k)$ to $i_{l,s}^{\mathrm{max}}= \max_k i_{l,s}(k)$. As a result, $\boldsymbol{\Xi}_{\mathrm{A}}$ is characterized as a row-wise block matrix
	\begin{equation}
		\boldsymbol{\Xi}_{l,\mathrm{A}} = \left[\mathbf{0}_{K\times (I_{l,e}-1)}^{\mathrm{T}}, \mathbf{U}_{K \times (I_{l,s}-I_{l,e}+1)}^{\mathrm{T}}, \mathbf{0}_{K\times (N-I_{l,s})}^{\mathrm{T}} \right]^{\mathrm{T}},
		\label{block1}
	\end{equation} 
	where $\mathbf{U}$ is a non-zero matrix; index $I_{l,e}$ and $I_{l,s}$ are given by $I_{l,e} = \lceil {(i_{l,e}^{\mathrm{min}}+1)N}/{2} \rceil$ and $I_{l,s} = \lceil {(i_{l,s}^{\mathrm{max}}+1)N}/{2} \rceil$.
	
	Further, we examine the IDFT of $\boldsymbol{\Theta}(\varphi_l,\psi_l)$ in the delay domain, i.e., $\boldsymbol{\Xi}_{l,\mathrm{D}}= \boldsymbol{\Theta}(\varphi_l,\psi_l)\mathbf{F}_{\mathrm{D}}^{*}$, where $\mathbf{F}_{\mathrm{D}}$ denote a $K$-dimension DFT matrix. Specifically, we have 
	\begin{equation}
		\begin{aligned}
			[\boldsymbol{\Xi}_{l,\mathrm{D}}]_{n,m}
			&\propto\frac{1}{\sqrt{K}}\sum_{k=0}^{K-1} \mathrm{e}^{\mathrm{j}\frac{2\pi}{K}mk} \mathrm{e}^{-\mathrm{j}2\pi k t_{l,n}}\\
			&\propto\frac{1}{\sqrt{K}} \frac{\sin\left(\pi K(t_{l,n} - \frac{m}{K})\right)}{\sin\left(\pi (t_{l,n} - \frac{m}{K})\right)},
		\end{aligned}
	\end{equation} 
	where $t_{l,n} = n^2\frac{f_s\varphi_l}{Kf_c} - n\frac{f_s\psi_l}{Kf_c}$, which is a quadratic function about $n$. Further, we have 
	\begin{equation}
		\lim_{K\to \infty} \left|[\boldsymbol{\Xi}_{l,D}]_{n,m}\right| \propto \sqrt{K}\delta\left(t_{l,n}-\frac{m}{K}\right).
		\label{delta}
	\end{equation}
	
	From (\ref{delta}), when $K\to \infty$, $[\boldsymbol{\Xi}_{l,D}]_{n,:}$ can be approximated as a delta function centered at $m = t_{l,n} K$. However, due to the finite sampling size $K$ in  practice, we consider the range from $j_{l,s}(n) = (t_n - {2}/{K}+1/2)K$ to $j_{l,e}(n) =(t_n + {2}/{K}+1/2)K$ to collect the delay indices near the peak effectively. 
	Extending this observation across all $N$ antennas, $\boldsymbol{\Xi}_\mathrm{D}$ is characterized as a column-wise block matrix
	\begin{equation}
		\boldsymbol{\Xi}_{l,\mathrm{D}} = \left[\mathbf{0}_{N\times (J_{l,s}-1)}, \mathbf{L}_{N \times (J_{l,e}-J_{l,s}+1)}, \mathbf{0}_{N\times (K-J_{l,e})} \right],
		\label{block2}
	\end{equation} 
	where  $J_{l,s} = \min_n j_{l,s}(n)$ and $J_{l,e} = \max_n j_{l,e}(n)$ with $n\in \boldsymbol{\phi}_l$, and $\mathbf{L}$ is a non-zero matrix.
	
	According to row-wise and column-wise block sparsity of (\ref{block1}) and (\ref{block2}), we assert that $\boldsymbol{\Xi}_l = \mathbf{F}_{\mathrm{A}}^{\mathrm{H}}\boldsymbol{\Theta}(\varphi_l,\psi_l)\mathbf{F}_{\mathrm{D}}^{\mathrm{*}}$ adheres to the following block-sparse structure, with its significant entries localized within the square region $\mathcal{A}_{l} \triangleq \left\{(n,k) \in \mathbb{Z}^{2} \mid I_{l,e} \le n \le I_{l,s}, J_{l,e} \le k \le J_{l,s} \right\}$, i.e., 
	\begin{equation}
		\boldsymbol{\Xi}_l = \left[\begin{array}{c:c:c}
			\mathbf{0} & \mathbf{0} & \mathbf{0}\\
			\hdashline
			\mathbf{0} & \mathbf{V}_{(I_{l,s}-I_{l,e}+1)\times(J_{l,e}-J_{l,s}+1)} & \mathbf{0}\\
			\hdashline
			\mathbf{0} & \mathbf{0} & \mathbf{0}\\ 
		\end{array}\right],
	\end{equation}
	where $\mathbf{V}$ denotes a non-zero matrix.
	
	Moreover, for a fixed column index $k$, $\boldsymbol{\Xi}_l$ characterizes the angular distribution corresponding to a specific propagation delay, which is determined by the spatial frequencies across different subcarriers. 
	Accordingly, within the region $\mathcal{A}_{l}$, each column captures the angular response for a given delay, and the width of significant spatial frequency components is approximately determined by the system bandwidth and the start spatial frequency, i.e., $i_{l,k} N {f_s} / {f_c}$, where $i_{l,k} \in [i_{l,e}^{\min}, i_{l,s}^{\max}]$ denotes the starting spatial frequency of the $k$-th column. 
	As a result, the nonzero components of $\boldsymbol{\Xi}_l$ are confined within a narrow band in each column, leading to a distinct column-wise clustered sparsity pattern.
	
	\section*{Appendix B A Proof of Theorem 1}
	\setcounter{equation}{0}
	\renewcommand\theequation{B.\arabic{equation}} 
	Now, we examine the 2D-IDFT of channel $\mathbf{H}$, i.e., $\mathbf{X} = \mathbf{F}_{\mathrm{A}}^{\mathrm{H}}\mathbf{H}\mathbf{F}_{\mathrm{D}}^{\mathrm{*}}$. By leveraging the linearity of the 2D-IDFT, we can focus solely on analyzing the $l$-th path. According to (\ref{H_matrix2}), we further define 
	\begin{equation}
		\mathbf{H}_{l,p} = \alpha_lc_{l,p}
		\left(\mathbf{b}_{\mathrm{far}}(\epsilon_{l,p}) \mathbf{a}^{\mathrm{T}}(\tau_l)\right)\odot\boldsymbol{\Theta}(\psi_l, \varphi_l).
	\end{equation}
	Applying the shift property of 2D-IDFT, $\mathbf{X}_{l,p} = \mathbf{F}_{\mathrm{A}}^{\mathrm{H}}\mathbf{H}_{l,p}\mathbf{F}_{\mathrm{D}}^{\mathrm{*}}$ should satisfy the block sparsity with its significant entries localized within the square region $\mathcal{B}_{l,p}$ with 
	\begin{equation}
		\begin{aligned}
			\mathcal{B}_{l,p} &= \left\{(n,k)\in\mathbb{Z}^2 \mid n = \mathrm{mod}(n_{l,1} + N \epsilon_{l,p}, N), \right.\\ 
			k &\left.= \mathrm{mod}(k_{l,1} + f_s\tau_l, K), \forall (n_{l,1}, k_{l,1}) \in \mathcal{A}_{l}\right\}.
		\end{aligned}
	\end{equation}
	Here, $\mathcal{A}_{l}$ is the non-zero region for  $\boldsymbol{\Xi}_l$ given by (\ref{region1}), and {mod}$(a, m)$ is the modulus of $a$ for $m$. 
	Considering all $P$ frequency components, the significant entries of $\mathbf{X}_l$ would be localized within the region $\mathcal{B}_{l} \triangleq \cup_p \mathcal{B}_{l,p}$. Thus, the proof is completed.
	\bibliographystyle{IEEEtran}
	\vspace{-1em}
	\bibliography{ref}
\end{document}